\documentclass[11pt,amssymb,epsfig,amsmath]{article}

\usepackage{graphicx,epsfig}
\usepackage[english]{babel}
\usepackage{amssymb,amsfonts,amsmath}

\newcommand{\be}{\begin{equation}}
\newcommand{\ee}{\end{equation}}
\newcommand{\ba}{\begin{eqnarray}}
\newcommand{\ea}{\end{eqnarray}}
\newcommand{\no}{\nonumber \\}
\newcommand{\gsim}{\mathrel{\hbox{\rlap{\lower.55ex \hbox {$\sim$}}
                   \kern-.3em \raise.4ex \hbox{$>$}}}}
\newcommand{\lsim}{\mathrel{\hbox{\rlap{\lower.55ex \hbox {$\sim$}}
                   \kern-.3em \raise.4ex \hbox{$<$}}}}

\def\del{{\partial}}

\def\roughly#1{\mathrel{\raise.3ex\hbox{$#1$\kern-.75em%
\lower1ex\hbox{$\sim$}}}}
\def\lsim{\roughly<}
\def\gsim{\roughly>}

\def\({\left(}
\def\){\right)}
\def\[{\left[}
\def\]{\right]}

\def\lsim{\mathrel{\rlap{\lower3pt\hbox{\hskip1pt$\sim$}}
     \raise1pt\hbox{$<$}}} 
\def\gsim{\mathrel{\rlap{\lower3pt\hbox{\hskip1pt$\sim$}}
     \raise1pt\hbox{$>$}}} 
\def\N{${\cal N}\,\,$}

\def\ka{\kappa}

\def\dlt{\delta}

\def\eps{\epsilon}
\def\omg{\omega}

\newcommand{\Gm}{\Gamma}

\newcommand{\tb}{{\bar t}}

\newcommand{\deq}{\Relbar\hspace{-0.3cm}{}^d}
\newcommand{\pht}{{\tilde \phi}}
\newcommand{\Z}{\int_0^z\frac{dz'}{f(z')}}
\newcommand{\tk}{t_K}
\newcommand{\xk}{x_K}

\begin{document}

\renewcommand{\thefootnote}{\arabic{footnote}}
\setcounter{footnote}{0}

\begin{flushright}
MPP-2012-98
\end{flushright}
\hfill {\today} \vskip 1cm


\begin{center}
{\LARGE\bf Thermalization from gauge/gravity duality: 

Evolution of singularities in unequal time correlators}
\date{\today}

\vskip 1cm {
\large Johanna Erdmenger\footnote{E-mail: jke@mpp.mpg.de} and
Shu Lin\footnote{E-mail: slin@mpp.mpg.de}\\
Max-Planck-Institut f\"{u}r Physik (Werner-Heisenberg-Institut) 
\\ F\"{o}hringer Ring 6, 80805 M\"{u}nchen, Germany
 }


\end{center}

\vskip 0.5cm

\begin{center}

\end{center}

\vskip 0.5cm

\begin{abstract}

We consider a gauge/gravity dual model of thermalization which
consists of a collapsing thin matter shell in asymptotically Anti-de
Sitter space. A central aspect of our model is to consider a shell
moving at finite velocity as determined by its equation of motion,
rather than a quasi-static approximation as considered previously in
the literature. By applying a divergence matching method, we obtain
the evolution of singularities in the retarded unequal time correlator
$G^R(t,t')$, which probes different stages of the thermalization. We
find that the number of singularities decreases from a finite number
to zero as the gauge theory thermalizes. This may be interpreted as a
sign of decoherence. Moreover, in a second part of the paper, we show
explicitly that the thermal correlator is characterized by the
existence of singularities in the complex time plane. 
By studying a quasi-static state, we show the singularities at real
times originate from contributions of normal modes. We also
investigate the possibility of obtaining complex singularities from
contributions of quasi-normal modes.

\end{abstract}

\newpage

\renewcommand{\thefootnote}{\#\arabic{footnote}}
\setcounter{footnote}{0}




\section{Introduction and Summary}

Recent experimental results have shown that
collective phenomena observed in heavy ion collisions at the
relativistic heavy ion collider (RHIC) are well-described by
hydrodynamics. The use of hydrodynamics assumes
a short  thermalization time, $\tau\sim0.5 fm$, which is the time
scale for the matter produced in the collisions to reach local
equilibrium.  The new experiments using
heavy ion collisions at the Large Hadron Collider (LHC) are expected
to provide further constraints on the thermalization time. The short
thermalization time is believed to be due to the strongly coupled
nature of the matter produced in the collisions. A theoretical
understanding of the thermalization mechanism requires knowledge of
Quantum Chromodynamics (QCD) dynamics at strong coupling and far from
equilibrium, which is inaccessible to lattice simulations and to 
perturbative field theory techniques. Gauge/gravity duality offers a
useful tool to study the dynamics of strongly coupled gauge
theory. Within the framework of gauge/gravity duality, the gravity
dual of $\N=4$ Super Yang-Mills theory at finite temperature is given by 
the AdS-Schwarzschild black hole geometry. It is therefore natural to
conjecture that the thermalization process is dual to black hole
formation via gravitational collapse in Anti-de Sitter space. 
Solving the necessary Einstein equations generally
requires numerical computations. Work along this line includes
\cite{minwalla,CY,bizon,zayas,heller,gubser,star}. 
While these works provide valuable information on the evolution of
the one-point function in the gauge theory undergoing thermalization,
perhaps equally important is the evolution of the two-point function,
which encodes information on the correlation and spectrum of a given
operator. The study of the two-point function amounts to a study of the
behavior of a bulk field in a gravitational collapse
background. Recently, there have been intensive efforts in studying
the evolution of the two-point function within gauge/gravity
duality. An incomplete list in the context of thermalization in heavy
ion collisions can be found in
\cite{uppsala,giddings,HLR,shell,kovchegov,ELN,EHL,11author,teaney,headrick,lopez,johnson,takayanagi}. In
particular, equal-time quantities have been extensively studied in
\cite{11author}, see also \cite{takayanagi}, which showed interesting
patterns of spatial correlation. On the other hand, unequal-time
correlators, in which the operators are inserted at different times,
allow to probe the causal structure and temporal correlation 
of the gauge theory and are therefore complementary to equal-time
quantities. 
In \cite{teaney}, an initial value problem was formulated for the unequal-time correlator and was applied to the background of an evolving black hole \cite{teaney2}. 

In this paper, we study the gravitational collapse of a massive
shell as in \cite{shell}. We include the finite velocity motion of the
shell in our analysis, thus going beyond the quasi-static
approximation used previously. The study of a massive shell has the advantage that the
corresponding field theory correlators have manifest singularities.
This feature is absent in other collapsing shell models, for instance
for the Vaidya metric, where the shell is
massless. For studying the singularities in the gravitational collapse
model, we use techniques which we developed in our previous
papers, where we considered the toy model of a mirror moving in
Anti-de Sitter space. In \cite{ELN}, we
have calculated the spatially-integrated unequal time correlator by
studying a bulk scalar in a moving ``mirror'' background, that is the
scalar satisfies the Dirichlet boundary condition on a prescribed
surface. We found that in the WKB limit, the singularities of the
resulting correlator showed a pattern consistent with a geometric
optics picture: The singularities occur at those times a light ray
bouncing between the AdS boundary and shell is 
reflected at the boundary. For a static mirror, this was already
realized in \cite{hoyos}.
The relation between singularity locations and geometric optics
can be viewed as a realization of the bulk-cone
singularity conjecture formulated in \cite{HLR}.  In a follow-up paper \cite{EHL}, the
present authors and Hoyos developed a powerful divergence matching
method, which allows for the determination of the precise form of the
singularities without solving the equation of motion for the scalar in
the bulk. This has paved the road for studying of the structure of the
singularities in a gravitational collapse model.

In this paper, we generalize the divergence matching method in order
to apply it to the case of gravitational collapse. We 
focus on the singularities of the retarded correlator
$G_R(t,t')=\theta(t-t') \langle [O(t),O(t')] \rangle$. For  varying
$t'$, we analyze the singularities of the correlator $G_R(t,t')$ 
and find that  singularities occur at $t=\tb_n$, where
$\tb_n$ ($\tb_1<\tb_2<\cdots$) is the time for a light ray originally leaving the
boundary at $t'$ to return to the boundary after the $n$-th bouncing
off the shell. The main result we find 
using the divergence matching method is an evolution of the
singularities in the process of thermalization. We will see that
unlike in the problem of a moving ``mirror'' background, where there
are infinite number of singularities in the correlator, the case of
gravitational collapse only contains a finite number of
singularities. In particular, the $n$-th singularity of $\tb_n$ moves
monotonously to $+\infty$ as $t'$ approaches a critical value $T_n$
($T_1>T_2>\cdots$) from below.  For $t'>T_1$, the last singularity
$\tb_1$ escapes from detection  and the correlator appears thermal as
far as the singularities are concerned. Since the singularities in the
unequal time correlator measure the strongest correlation in time, the disappearance of singularities has the interpretation of temporal decoherence.

A second aspect we study in this paper is the correlator for a field
in thermal equilibrium, which  is characterized by the appearance of
singularities in the complex $t$ plane, which is closely related to
the structure of the quasi-normal modes (QNM). The geometric optics
picture has been used in \cite{hoyos} to extract the asymptotic QNM
in the AdS-Schwarzschild background, where the asymptotic QNM are
obtained as the reciprocal of the complex time period of a light ray
bouncing in AdS-Schwarzschild background. We confirm this picture by
explicitly evaluating the retarded correlator and identifying the
singularities in the complex $t$ plane. To understand the appearance
of the singularities in the complex $t$ plane, we further investigate
the evolution of the QNM for the quasi-static states  dual
to a shell levitating at rest at different positions above the horizon. We study the evolution of the QNM as the shell is lowered to the horizon. We find that among the QNM the normal modes allows us to reproduce the singularities for real $t$ obtained by the divergence matching method, while the complex QNM does not lead to the expected singularities in the complex $t$ plane.

The paper is structured as follows: In Section 2, we develop a
differential form of the divergence matching method, which is more
suitable for generalization to the gravitational collapse model. In
Section 3, we review the gravitational collapse model used in
\cite{shell} and study the behavior of a bouncing light ray in the
collapsing background. Then we generalize the divergence matching
method to the case of gravitational collapse model and check that it
passes a non-trivial test. The application of the method gives rise to
the singularity evolution described above. Section 4 is devoted to the
calculation of the unequal-time correlator in thermal equilibrium,
which shows the appearance of the singularity in the complex time
plane. In Section 5, we revisit the results of Section 3 and give a
dual picture of the evolution of the QNM with an explicit example of a
sequence of quasi-static states.
We discuss open questions and future directions in Section 6. Some
notes on the computation of black hole QNM are
 collected in the appendix.

\section{Differential form of divergence matching for Dirichlet problem}

Let us first reformulate the divergence matching method developed in \cite{EHL} for the Dirichlet problem in a differential form, which is easily generalizable to more complicated models. We start with the Dirichlet problem
\begin{align}\label{Gr_diff}
\left\{\begin{array}{l}
\square G^R(t,z,t')=0\\
G^R(t,z\to0,t')=\dlt(t-t')\\
G^R(t,z=f(t),t')=0
\end{array}
\right..
\end{align}
Here $z=f(t)$ is the Dirichlet surface, which can be viewed as a mirror. Without the mirror,\begin{align} 
G^R(t,z,t')=G_0^R(t-t',z)=B_0\sum_{+,-}\frac{\pm z^d}{(-(t-t'\mp i\eps)^2+z^2)^c}\theta(t-t'),
\end{align}
where
\begin{gather}\label{c}
c={\frac{d+1}{2}} \, , \qquad
B_0=\frac{i}{\pi}\frac{\Gm(c)\Gm(\frac{1}{2})}{\Gm(\frac{d}{2})2^c}.
\end{gather}
The presence of the mirror will scatter $G^R$ in a way consistent with its retarded nature. The net result we will find is that the divergences are propagated through the bouncing of $G^R$ between the mirror and the boundary.

\begin{figure}
\includegraphics[width=0.5\textwidth]{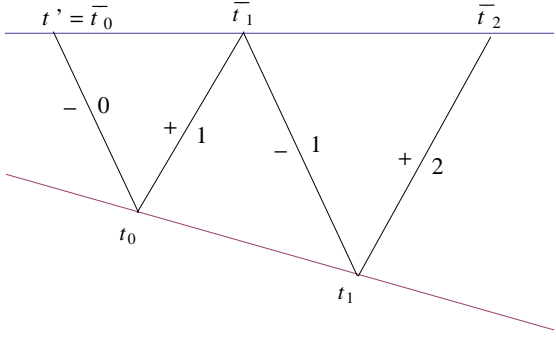}
\caption{\label{bouncing} The scalar field in the vicinity the segments labelled by $\mp$ contains divergence of ingoing/outgoing type. The segments of the trajectory are also labelled by $n$ if they are connected to the boundary point $\tb_n$. See the text for more details.}
\end{figure}

Fig.\ref{bouncing} illustrates the mechanism of the divergence
matching method: A light ray starting at $t'$ from the boundary
follows a bouncing trajectory composed of segments of null
geodesics. We label each segment by a number $n$ and sign $-/+$. The
number label $n$ indicates that the segment is connected to the
boundary point $\tb_n$. 
Two segments joining at the same boundary point have the obvious
interpretation as corresponding to 
the ingoing and outgoing waves, respectively, which we label by $-$
and $+$.  In the vicinity of each segment, $G^R$ is singular. In the
following, we will simply refer to the  divergences of $G^R$ as being
of ingoing and outgoing type for the sign label $-$ or $+$,
respectively. The idea is to determine all the  divergences along
$(n,+)$, $(n,-)$ through matching along  the mirror trajectory and on
the boundary.  The matching procedure is explained in detail in the subsequent.

The initial data for the divergence along segment $(0,-)$ is provided by $G_0^R$, just as in \cite{EHL}. It is given by
\begin{align}\label{G0_diff}
G_0^-=\frac{G_0^R(t-t'\to z)}{z^{\frac{d-1}{2}}}=\sum_{+,-}\frac{\pm B_0}{(-t+t'\pm i\eps+z)^c},
\end{align}
with $c$ as in \eqref{c}.
Bearing in mind that the prescription $t\to t-i\eps$ ($t\to t+i\eps$)
corresponds to a positive (negative) frequency contribution, we may identify the upper/lower signs as contributions from positive/negative frequencies. We may treat them separately in what follows. Using $>$/$<$ for positive/negative frequencies, we write
\begin{align}\label{G0_pm}
G_{0,>}^-=\frac{B_0}{(-t+t'+i\eps+z)^c},\quad\quad G_{0,<}^-=\frac{-B_0}{(-t+t'-i\eps+z)^c}.
\end{align}
The first step of matching is to be done in the vicinity of $(t_0,z_0=f(t_0))$. Along segment $(1,+)$, the outgoing type divergence of $G^R$ assumes the generic form
\begin{align}\label{G1_pm}
G_{1,>}^+=\frac{B_1}{(\tb_1-t-z+i\eps)^c},\quad\quad G_{1,<}^+=\frac{-B_1}{(\tb_1-t-z-i\eps)^c}.
\end{align}
In the vicinity of $(t_0,f(t_0))$, the divergence of $G^R$ is a superposition of $G_0^-$ and $G_1^+$. The Dirichlet boundary condition on $z=f(t)$ gives
\begin{align}\label{Dirichlet_pm}
G_{0,>}^-+G_{1,>}^+=0,\quad\quad G_{0,<}^-+G_{1,<}^+=0.
\end{align}
Denoting $t=t_0(1+x)$, then $z=f(t_0)+f'(t_0)t_0x$. Plugging \eqref{G0_pm} and \eqref{G1_pm} into \eqref{Dirichlet_pm} and expanding to leading order in $x$, we obtain
\begin{align}
B_1=-B_0\(\frac{1+f'(t_0)}{1-f'(t_0)}\)^c.
\end{align}
The next step is to do the matching at the boundary close to $\tb_1$,
where $G^R$ is a superposition of $G_1^+$ and $G_1^-$. This will allow
us to determine $G_{1}^-$. It is convenient to use the frequency
representation. We consider the spatially integrated correlator which
corresponds to the zero momentum mode. Therefore, for a single
component of frequency we have, assuming $\omg>0$,
\begin{align}\label{freq_rep_d}
&G_>^-:\;z^{\frac{d}{2}}e^{-i\omg t}H_{\frac{d}{2}}^{(1)}(\omg z)=
z^{\frac{d-1}{2}}\sqrt{\frac{2}{\pi\omg}}e^{-i\omg t+i(\omg
  z-\frac{\pi c}{2})}\, , \quad \omg\to\infty \, , \\
&G_<^-:\;z^{\frac{d}{2}}e^{-i\omg t}H_{\frac{d}{2}}^{(2)}(-\omg z)=
z^{\frac{d-1}{2}}\sqrt{\frac{2}{-\pi\omg}}e^{-i\omg t+i(\omg
  z+\frac{\pi c}{2})} \, , \quad \omg\to\infty, \\
&G_>^+:\;z^{\frac{d}{2}}e^{-i\omg t}H_{\frac{d}{2}}^{(2)}(\omg z)=
z^{\frac{d-1}{2}}\sqrt{\frac{2}{\pi\omg}}e^{-i\omg t-i(\omg
  z-\frac{\pi c}{2})}\, , \quad \omg\to\infty \, , \\
&G_<^+:\;z^{\frac{d}{2}}e^{-i\omg t}H_{\frac{d}{2}}^{(1)}(-\omg z)=
z^{\frac{d-1}{2}}\sqrt{\frac{2}{-\pi\omg}}e^{-i\omg t-i(\omg
  z+\frac{\pi c}{2})}\, , \quad  \omg\to\infty.
\end{align}
In order to make the boundary free of source away from $t'$, we need
to require the superposition of $G_1^+$ and $G_1^-$ contains
$z^{\frac{d}{2}}J_{\frac{d}{2}}(|\omg|z)$ only, thus for each
$H_{\frac{d}{2}}^{(1)}(\omg z)$ in $G_>^-$, we need
 $H_{\frac{d}{2}}^{(2)}(\omg z)$ in $G_>^+$. Similarly, for each
 $H_{\frac{d}{2}}^{(2)}(-\omg z)$ in $G_<^-$, we need
 $H_{\frac{d}{2}}^{(1)}(-\omg z)$ in $G_<^+$. Using the representation \eqref{freq_rep_d}, we conclude that
\begin{align}
G_{1,>}^-=\frac{B_1e^{-i\pi c}}{(\tb_1-t+z+i\eps)^c},\quad G_{1,<}^-=\frac{-B_1e^{i\pi c}}{(\tb_1-t+z-i\eps)^c}.
\end{align}
Repeating this process, we find the divergences propagated along the null geodesics are given by
\begin{align}\label{Gn_diff}
&G_{n,>}^+=\frac{B_n(e^{-i\pi c})^{n-1}}{(\tb_n-t-z+i\eps)^c},\quad G_{n,<}^+=\frac{-B_n(e^{i\pi c})^{n-1}}{(\tb_n-t-z-i\eps)^c}, \no
&G_{n,>}^-=\frac{B_n(e^{-i\pi c})^n}{(\tb_n-t-z+i\eps)^c},\quad G_{n,<}^-=\frac{-B_n(e^{i\pi c})^n}{(\tb_n-t-z-i\eps)^c},
\end{align}
where $B_n$ satisfies the recursion relation $B_n=-B_{n-1}\(\frac{1+f'(t_{n-1})}{1-f'(t_{n-1})}\)^c$.
Finally, we wish to extract the divergences of the unequal-time correlator again by taking advantage of the frequency representation, for which we also need the coefficients of the $z^d$ term in the expansion near $z=0$,
\begin{align}
&G_>^-:\;z^{\frac{d}{2}}e^{-i\omg t}H_{\frac{d}{2}}^{(1)}(\omg z)\to
\frac{-e^{\frac{-i\pi d}{2}}}{i\sin\frac{\pi
    d}{2}}\frac{1}{\Gm(\frac{d}{2}+1)}\(\frac{\omg}{2}\)^{\frac{d}{2}}e^{-i\omg
  t}z^d \, , \;\;\quad \\
&G_<^-:\;z^{\frac{d}{2}}e^{-i\omg t}H_{\frac{d}{2}}^{(2)}(-\omg z)\to
\frac{e^{\frac{i\pi d}{2}}}{i\sin\frac{\pi
    d}{2}}\frac{1}{\Gm(\frac{d}{2}+1)}\(\frac{-\omg}{2}\)^{\frac{d}{2}}e^{-i\omg
  t}z^d \, , \; \\
&G_>^+:\;z^{\frac{d}{2}}e^{-i\omg t}H_{\frac{d}{2}}^{(2)}(\omg z)\to
\frac{e^{\frac{i\pi d}{2}}}{i\sin\frac{\pi
    d}{2}}\frac{1}{\Gm(\frac{d}{2}+1)}\(\frac{\omg}{2}\)^{\frac{d}{2}}e^{-i\omg
  t}z^d \, , \;\;\quad  \\
&G_<^+:\;z^{\frac{d}{2}}e^{-i\omg t}H_{\frac{d}{2}}^{(1)}(-\omg z)\to
\frac{-e^{\frac{-i\pi d}{2}}}{i\sin\frac{\pi d}{2}}\frac{1}{\Gm(\frac{d}{2}+1)}\(\frac{-\omg}{2}\)^{\frac{d}{2}}e^{-i\omg t}z^d\; .
\end{align}
$G_{n,>}^+$, $G_{n,<}^+$, $G_{n,>}^-$ and $G_{n,<}^-$ lead to four contributions to the divergence of $G^R(t\to\tb_n,t')$. We calculate them one by one,
\begin{align}\label{Gn>-}
G_{n,>}^-=\frac{B_ne^{-in\pi c}}{(\tb_n-t+z+i\eps)^c}=\int_0^\infty d\omg g_>(\omg)\sqrt{\frac{2}{\pi\omg}}e^{-\frac{i\pi c}{2}}e^{-i\omg(t-z-i\eps)},
\end{align}
where $g_>(\omg)$ is the weight for frequency $\omg$.
Thanks to the $i\eps$ prescription, the above can be identified as Laplace transform, which can be inverted to give
\begin{align}
g_>(\omg)=\frac{\sqrt{\frac{\pi\omg}{2}}B_ne^{-in\pi c}e^{i\omg\tb_n}\omg^{c-1}}{\Gm(c)}.
\end{align}
The contribution to the divergence of $G^R(t\to\tb_n,t')$ is then given by
\begin{align}
&\int_0^\infty d\omg g_>(\omg)\frac{-e^{\frac{-i\pi d}{2}}}{i\sin\frac{\pi d}{2}}\frac{1}{\Gm(\frac{d}{2}+1)}\(\frac{\omg}{2}\)^{\frac{d}{2}}e^{-i\omg(t-i\eps)} \no
 \\ & \hspace{1cm} = \, -\frac{\sqrt{\frac{\pi}{2}}B_n}{\Gm(c)}\frac{e^{-\frac{i\pi d}{2}}e^{-in\pi c}}{i\sin\frac{\pi d}{2}}\frac{1}{\Gm(\frac{d}{2}+1)}\(\frac{1}{2}\)^{\frac{d}{2}}\frac{\Gm(d+1)}{(-i\tb_n+it+\eps)^{2c}}.
\end{align}
Similarly,
\begin{align}\label{Gn<-}
&G_{n,<}^- \, = \, \frac{-B_ne^{in\pi
    c}}{(\tb_n-t+z-i\eps)^c}=\int_{-\infty}^0 \! d\omg \,
g_<(\omg)\sqrt{\frac{2}{-\pi\omg}}e^{\frac{i\pi c}{2}
}e^{-i\omg(t-z+i\eps)} \, , \end{align}
which implies
\begin{align}
&g_<(-\omg) \, = \, \frac{-\sqrt{\frac{\pi\omg}{2}}B_ne^{in\pi c}e^{i\omg\tb_n}\omg^{c-1}}{\Gm(c)}.
\end{align}
The contribution to the divergence of $G^R(t\to\tb_n,t')$ is then given by
\begin{align}
&\int_0^\infty d \! \omg \,  g_<(-\omg)\frac{e^{\frac{i\pi d}{2}}}{i\sin\frac{\pi d}{2}}\frac{1}{\Gm(\frac{d}{2}+1)}\(\frac{\omg}{2}\)^{\frac{d}{2}}e^{i\omg(t+i\eps)} \no
& \hspace{1cm} \, = \, -\frac{\sqrt{\frac{\pi}{2}}B_n}{\Gm(c)}\frac{e^{\frac{i\pi d}{2}}e^{in\pi c}}{i\sin\frac{\pi d}{2}}\frac{1}{\Gm(\frac{d}{2}+1)}\(\frac{1}{2}\)^{\frac{d}{2}}\frac{\Gm(d+1)}{(i\tb_n-it+\eps)^{2c}}.
\end{align}
The calculations of the contributions from $G_{n,>}^+$ and $G_{n,<}^+$ exactly parallel those for $G_{n,>}^+$ and $G_{n,<}^+$. We obtain
\begin{align}
&G_{n,>}^+:\;\frac{\sqrt{\frac{\pi}{2}}B_n}{\Gm(c)}\frac{e^{\frac{i\pi
      d}{2}}e^{-in\pi c}}{i\sin\frac{\pi
    d}{2}}\frac{1}{\Gm(\frac{d}{2}+1)}\(\frac{1}{2}\)^{\frac{d}{2}}\frac{\Gm(d+1)}{(-i\tb_n+it+\eps)^{2c}}
\, ,\\
&G_{n,<}^+:\;\frac{\sqrt{\frac{\pi}{2}}B_n}{\Gm(c)}\frac{e^{-\frac{i\pi d}{2}}e^{in\pi c}}{i\sin\frac{\pi d}{2}}\frac{1}{\Gm(\frac{d}{2}+1)}\(\frac{1}{2}\)^{\frac{d}{2}}\frac{\Gm(d+1)}{(i\tb_n-it+\eps)^{2c}}.
\end{align}
Summing over the four contributions, we obtain a neat result for the most singular part of $G_R(t,t')$ as $t\to\tb_n$:
\begin{align}\label{Gr_sum}
G_R(t\to\tb_n,t')=\frac{\sqrt{2\pi}\Gm(d+1)B_n}{2^{\frac{d}{2}}\Gm(c)\Gm(\frac{d}{2}+1)}\(\frac{e^{-i\pi c(n-1)}}{(-t+\tb_n+i\eps)^{2c}}-\frac{e^{i\pi c(n-1)}}{(-t+\tb_n-i\eps)^{2c}}\).
\end{align}
Comparing \eqref{Gr_sum} with the result obtained using 
 the integral form of the divergence matching method in \cite{EHL},  we find perfect agreement.

\section{Divergence matching for a gravitational collapse model}

\subsection{A gravitational collapsing shell}

In this section, we study a gravitational collapse model with the
purpose of gaining further insight into
thermalization of the dual gauge theory. The model has been described
in detail in \cite{shell} in the quasi-static approximation. We recall
the key ingredients here: The model contains a homogeneous shell
collapsing under its own gravity. The shell separates the spacetime
into the parts above and below. By above and below, we refer to the
region between AdS boundary and the shell and the region between the
shell and the AdS interior, respectively. The corresponding metrics
are given by AdS$_5$-Schwarzschild and by pure AdS,
\begin{align}\label{metrics}
\text{above}&: ds^2=\frac{-fdt_f^2+d{\vec x}^2+dz^2/f}{z^2} \\
\text{below}&: ds^2=\frac{-dt^2+d{\vec x}^2+dz^2}{z^2}, \\
\text{with}&\;\; f=1-\frac{z^4}{z_h^4}.
\end{align}
The $z_h$ is the position of horizon, which defines a temperature $z_h=\frac{1}{\pi T}$. As we will see soon, in Schwarzschild coordinates, the horizon is always ``protected'' by the shell, but in Kruskal coordinates, the shell will be able to cross the horizon. Note that we have used $t_f$ for the time coordinate above the shell to distinguish it from the coordinate $t$ below. However we choose the radial coordinate $z$ to be continuous across the shell. We choose to parameterize the hypersurface $\Sigma$ traced out by the shell by $\tau,\vec{x}$ such that the induced metric on $\Sigma$ is given by
\begin{align}\label{ind_metric}
ds_{\Sigma}^2=\frac{-d\tau^2+d\vec{x}^2}{z(\tau)^2}.
\end{align}
The trajectory of the shell as measured by the coordinates above the shell is $t_f(\tau),\,z(\tau)$. Similarly below the shell it is described by $t(\tau),\,z(\tau)$. The velocity of the shell is defined as $u^\mu=\frac{dx^\mu}{d\tau}$:
\begin{align}
u_f^\mu&=(\dot{t}_f,\vec{0},\dot{z})\quad\text{above}, \no
u^\mu&=(\dot{t},\vec{0},\dot{z})\quad\text{below}.
\end{align}
Comparing \eqref{metrics} and \eqref{ind_metric}, we obtain the following relations:
\begin{align}
f\dot{t}_f{}^2-\frac{\dot{z}^2}{f}=1,\;\dot{t}^2-\dot{z}^2=1.
\end{align}
The unit vector normal to the hypersurface $\Sigma$ satisfies $n\cdot u=0,\;n^2=1$. It can be obtained easily as
\begin{align}
n_f^\mu&=(\frac{\dot{z}z}{f},\vec{0},\dot{t}_fzf)\quad \text{above} \,
,\no
n^\mu&=(\dot{z}z,\vec{0},\dot{t}z)\quad \text{below}.
\end{align}

The falling trajectory of the shell is determined by the Israel junction conditions \cite{israel}. We have used implicitly the continuity of the induced the metric in \eqref{ind_metric}. A further condition is given by
\begin{gather}\label{junc_cond}
[K_{ij}-\gamma_{ij}K]=\ka S_{ij} \, , \qquad
\{K_{ij}\}S^{ij}=0.
\end{gather}
We use Greek letters for spacetime coordinates and Latin letters for coordinates on $\Sigma$. $\gamma_{ij}$ is simply the induced metric and $K_{ij}\equiv n_\alpha\(\frac{\del^2x^\alpha}{\del\xi^i\del\xi^j}+\Gamma^\alpha_{\beta\gamma}\frac{\del x^\beta}{\del\xi^i}\frac{\del x^\gamma}{\del\xi^j}\)$ is the extrinsic curvature. The square (curly) bracket denotes the difference (sum) of the quantities above and below the shell. $S_{ij}$ is the stress tensor of the shell. To proceed, we use the ideal fluid type stress tensor: $S_{ij}=(\eps(z)+p(z))u_iu_j+p(z)\gamma_{ij}$. Note that we allow the energy density and pressure to depend on the radial coordinate.

It is straightforward to calculate the nonvanishing components of the extrinsic curvature above the shell as
\begin{align}\label{extrinsic}
&K_{\tau\tau}=\frac{d\sqrt{f+\dot{z}^2}}{zdz}-\frac{\sqrt{f+\dot{z}^2}}{z^2}
\, , 
&K_{x1x1}=K_{x2x2}=K_{x3x3}=\frac{\sqrt{f+\dot{z}^2}}{z^2}. 
\end{align}
The counterpart below the shell is readily obtained by taking $f=1$.
Applying  \eqref{junc_cond}, we obtain the following constraint on the energy density and pressure,
\begin{align}
\frac{d\eps}{3dz}=\frac{\eps+p}{z}.
\end{align}
We choose the equation of state to be $\eps=\frac{p}{a}$, which gives
$\eps\sim z^{3(a+1)}$. Inserting this  into \eqref{junc_cond} leads to the  relations
\begin{align}
\sqrt{1+\dot{z}^2}-\sqrt{f+\dot{z}^2} &\sim z^{3(a+1)} \, , \\
\sqrt{1+\dot{z}^2}+\sqrt{f+\dot{z}^2} & \sim z^{1-3a} \, .
\end{align}
To be specific, we fix $a=\frac{1}{3}$, which corresponds to a conformal equation of state, thus
\begin{align}
\sqrt{1+\dot{z}^2}-\sqrt{f+\dot{z}^2}=bz^4,
\end{align}
with $b$ a parameter. It is a short exercise to show that
\begin{gather}\label{ztdot}
\dot{z} =\sqrt{\frac{1}{4}\(bz^4+\frac{1}{bz_h^4}\)^2-1} \, , \quad
\dot{t}_f  =\frac{\sqrt{f+\dot{z}^2}}{f}=\frac{\frac{1}{bz_h^4}-bz^4}{2f}.
\end{gather}
With \eqref{ztdot}, we can easily solve for the trajectory of the falling shell in the coordinate above the shell. A convenient way to parameterize the trajectory is to use the initial radial position $z_s$, at which the shell starts falling with vanishing velocity, i.e. $\dot{z}=0$. This gives a relation among three parameters $b$, $z_s$ and $z_h$,
\begin{align}
bz_s^4+\frac{1}{bz_h^4}=2.
\end{align}
This has the physical meaning that the temperature $z_h$ is determined by the energy density $b$ and the intrinsic scale $z_s$. From now on, we will set $z_h=1$, with the understanding that all other quantities are measured in units of the temperature. Since $z_s<z_h=1$, we may easily convince ourselves that $\frac{1}{2}<b<1$.

\subsubsection{Bouncing light rays in the gravitational background}

\begin{figure}
\includegraphics[width=0.4\textwidth]{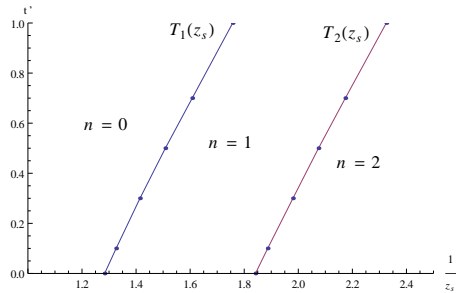}
\caption{\label{fig_n} The regions of parameter space marked by the number bouncings of the light ray on the shell. A smaller $t'$ and $z_s$ always lead to more bouncings. We observe a good linear relation between $\frac{1}{z_s}$ and $t'$ along the boundaries of regions $T_n$.}
\end{figure}

It is useful to study the behavior of light rays in the gravitational
collapse background before we compute the singular part of the
correlator in the next section. The massive shell always falls with a
speed less than that of light, thus naively the light ray will bounce
forever between the shell and the boundary. This is however not true
because of the warping, which freezes both the shell and the light at
the horizon. For a given parameter $z_s$ and the initial time $t'$
that the light leaves the boundary, there are only a finite number of
bouncings before the light ray asymptotes to the shell trajectory. The
problem of finding the bouncing trajectory of a light ray in the
gravitational collapse background can be studied with simple
numerics. Fig.\ref{fig_n} shows a chart of the number of bouncings in
the parameter space of intrinsic scale $z_s$ and the time parameter
$t'$. We have not mapped out the regions with $n>2$. The reason will
be clear later. For a given $z_s$, we may define $T_n(z_s)$ as the
critical value of $t'$, beyond which an $n$-th bouncing does not
occur. The $T_n$ are curves separating regions with different number
of bouncings. In general, smaller $t'$ and larger $\frac{1}{z_s}$ lead
to more bouncings. Physically, a small $t'$ means that the correlator
probes an earlier stage of the thermalization process,  while a larger
$\frac{1}{z_s}$ corresponds to a higher intrinsic scale the
thermalization process starts with. 
As argued in \cite{shell}, the scale should be related to the
saturation scale of the nucleus in the context of heavy ion collisions.

One may wonder if the above picture is an artefact of the
Schwarzschild coordinate. In fact, the picture is more transparent
 in Kruskal coordinates. Working with Kruskal coordinates defined by
\begin{gather}\label{kruskal}
e^{4r_*}=\xk^2-\tk^2 \, , \quad
\tanh2t=\frac{\tk}{\xk} \, , \quad
r_*=\int_z^\infty\frac{dz}{f}+\frac{i\pi}{4}  \, ,
\end{gather}
the AdS$_5$ Schwarzschild metric becomes
\begin{align}
ds^2=\frac{f}{4z^2}e^{-4r_*}\(-d\tk^2+d\xk^2\)+\frac{\vec x^2}{z^2}.
\end{align}
In terms of the Kruskal coordinates, the trajectory of the shell is determined from \eqref{ztdot} as
\begin{align}\label{tkxkdot}
&\dot\tk=\bigg[\(\frac{1}{b}-bz^4\)\xk-\sqrt{\(\frac{1}{b}+bz^4\)^2-4}\tk\bigg]\frac{1}{f} \no
&\dot\xk=\bigg[\(\frac{1}{b}-bz^4\)\tk-\sqrt{\(\frac{1}{b}+bz^4\)^2-4}\xk\bigg]\frac{1}{f}.
\end{align}
In Kruskal coordinates, the warp factor cancels when the shell reaches the falling velocity
\begin{align}\label{velocity_kruskal}
\frac{d\xk}{d\tk}=\frac{\(\frac{1}{b}-bz^4\)\tk-\sqrt{\(\frac{1}{b}+bz^4\)^2-4}\xk}{\(\frac{1}{b}-bz^4\)\xk-\sqrt{\(\frac{1}{b}+bz^4\)^2-4}\tk}.
\end{align}
It starts from $0$ initially at $z=z_s$ and asymptotes to $1$ as
$z\to\infty$. The shell will cross the horizon at finite Kruskal
time. Naively, at $z=1$,
$\frac{1}{b}-bz^4=\sqrt{\(\frac{1}{b}+bz^4\)^2-4}$, we may conclude
that $\frac{d\xk}{d\tk}=-1$ according to
\eqref{velocity_kruskal}. This is not true, however, because we also
have $\tk=\xk$. As a result \eqref{velocity_kruskal} becomes
undetermined, L'Hopital rule is needed in the evaluation. A more
careful analysis shows a positive velocity when the shell crosses the
horizon. Numerical integration of \eqref{velocity_kruskal} also
confirms that it is  a monotonous increasing function of $z$.

The light ray trajectory in Kruskal coordinates becomes trivial. It is
simply given by $\frac{d\xk}{d\tk}=\pm 1$, therefore we expect an  infinite number of bouncings between the shell and the boundary. For a given $z_s$, the critical value $T_n(z_s)$ corresponds to the situation that the $n$-th bouncing on the shell occurs precisely when the shell crosses the horizon. For $t'>T_n(z_s)$, the $n$-th bouncing occurs below the horizon, thus the reflected will not be able to escape from the point of view of the Schwarzschild coordinates.

\subsubsection{Divergence matching method for a probe scalar}

We now turn to the 
spatially integrated retarded correlator 
\begin{equation} 
G_R(t,t')=\int \! d^{d-1}x \, \theta(t-t') \, [O(t,\vec{x}),O(t',0)]
\end{equation}
 in the gravitational collapse background. We choose to study the
 behavior of a probe scalar in this background and extract the
 correlator of the dual operator $O(x)$. We will first derive the
 matching condition of the scalar on the shell, and then apply the
 method of divergence matching to determine the most singular part of
 the correlator without solving the scalar wave equation in the
 bulk. For the case of a scalar in the gravitational collapse
 background, the boundary condition on the hypersurface of the shell
 is more complicated, which requires a generalization of the
 divergence matching method formulated in the previous section. We
 will still work in general dimension $d$ , which  makes the structure more transparent. Finally we will set $d=4$.

The behavior of a scalar in the gravitational collapse background has been studied in \cite{giddings}. The matching condition is simply the continuity of the scalar itself and its flux. In our particular example, it is given by
\begin{align}\label{mat_cond}
\left\{\begin{array}{l}
\phi_f=\phi\\
n_f\cdot\nabla\phi_f=n\cdot\nabla\phi
\end{array}
\right..
\end{align}
The first line of \eqref{mat_cond} implies $u_f\cdot\nabla\phi_f=u\cdot\nabla\phi$, which combined with \eqref{mat_cond} gives the explicit relations
\begin{align}\label{mat_cond_ex}
\left\{\begin{array}{l}
\dot{t}_f\del_{t_f}\phi_f+\dot{z}\del_z\phi_f=\dot{t}\del_t\phi+\dot{z}\del_z\phi\\
\frac{\dot{z}}{f}\del_{t_f}\phi_f+\dot{t}_ff\del_z\phi_f=\dot{z}\del_t\phi+\dot{t}\del_z\phi \\
\phi_f=\phi
\end{array}
\right..
\end{align}
Since we are interested in the retarded correlator, the scalar wave
contains only the ingoing component below the shell. The wave in
frequency representation is $z^{\frac{d}{2}}H^{(1)}_{\frac{d}{2}}(\omg
z)$ or $z^{\frac{d}{2}}H^{(2)}_{\frac{d}{2}}(-\omg z)$ for positive
and negative frequency modes,  respectively. Working in the UV limit as in \cite{EHL}, we may use the asymptotic expansion of the Hankel function to show
\begin{align}\label{shell_vac}
\del_t\phi+\del_z\phi=\frac{d-1}{2z}\phi.
\end{align}
Plugging \eqref{shell_vac} into \eqref{mat_cond_ex}, we obtain a matching condition involving $\phi_f$ only,
\begin{align}\label{shell_th}
\(\dot{t}_f+\frac{\dot{z}}{f}\)\del_{t_f}\phi_f+(\dot{z}+\dot{t}_ff)\del_z\phi_f=(\dot{t}+\dot{z})\frac{d-1}{2z}\phi_f,
\end{align}
where $\dot{t}_f=\frac{\sqrt{f+\dot{z}^2}}{f}$ and $\dot{t}=\sqrt{1+\dot{z}^2}$. All the quantities are to be evaluated on the shell.

\subsection{Method of divergence matching}

Now we  derive a recursion relation for the most singular part of the correlator. We again refer to Fig.\ref{bouncing}. As in the Dirichlet problem in Section.2, close to each segment of the trajectory, the singular part of the scalar wave is either ingoing (for the minus sign) or outgoing (for the positive sign). Furthermore, working in the UV limit allows us to approximate the ingoing/outgoing waves by WKB solutions. From now on, we will suppress the subscript $f$ when the region below the shell does not enter the discussion. We  construct the WKB solutions
\begin{align}\label{WKB_sol}
\phi_\mp=\int d\omg g_\mp(\omg)z^{\frac{d-1}{2}}e^{\pm i\omg\int_0^z\frac{dz'}{f(z')}}e^{-i\omg t},
\end{align}
which have the  obvious property
\begin{align}\label{prop}
\(\pm\del_t+f\del_z\)\phi_\mp=\frac{d-1}{2z}\phi_\mp f.
\end{align}
We split the wave into ingoing and outgoing components $\phi_f=\phi_-+\phi_+$. Inserting this into \eqref{shell_th} and using \eqref{prop}, we obtain
\begin{align}\label{key_eq}
2\del_t\phi_+=\frac{f\(\sqrt{1+\dot{z}^2}-\sqrt{f+\dot{z}^2}\)}{\sqrt{f+\dot{z}^2}+\dot{z}}\frac{d-1}{2z}\(\phi_-+\phi_+\).
\end{align}
This is the key equation for divergence matching.

We begin with the matching at $(t_0,z_0)$. As we argued before, close
to the segment $0,-$, the singular part of the wave is ingoing
$\phi_{0,-}$, and close to $1,+$, the singular part of the wave is
outgoing $\phi_{1,+}$. There is an overlap region between the two,
which is close to $(t_0,z_0)$ on the shell. Since $\phi_{0,-}$ is
simply the WKB limit of thermal bulk-to-boundary correlator  with
ingoing component only, a WKB computation or simple analogue with zero
temperature bulk-to-boundary correlator gives
\begin{align}
\phi_{0,-}&=z^{\frac{d-1}{2}}B_0\bigg[\frac{1}{\(-t+t'+i\eps+\Z\)^c}-\frac{1}{\(-t+t'-i\eps+\Z\)^c}\bigg] \no
&=\phi_{0,-}^>+\phi_{0,-}^< , \end{align}
{with}
\begin{gather}
\phi_{0,-}^> =\frac{z^{\frac{d-1}{2}}B_0}{\(-t+t'+i\eps+\Z\)^c} \, , \quad
\phi_{0,-}^< =-\frac{z^{\frac{d-1}{2}}B_0}{\(-t+t'-i\eps+\Z\)^c} \, .
\end{gather}
The contributions with a $>$ and $<$ sign have a clear interpretation
according to the $i\eps$ prescription: They correspond to the contributions
from positive and negative frequency, respectively.
Since we are interested in the most singular part of $\phi_{1,+}$, we
may drop the term proportional to $\phi_{1,+}$ on the right hand side
of \eqref{key_eq}. It is easy to guess the general form of the singular part to be
\begin{align}
\phi_{1,+}^>&=\frac{z^{\frac{d-1}{2}}B_1}{\(-t+t'+i\eps+\Z\)^{c-1}} \,
, \no
\phi_{1,+}^<&=-\frac{z^{\frac{d-1}{2}}B_1}{\(-t+t'-i\eps+\Z\)^{c-1}}
\, .
\end{align} 
Inserting the above into \eqref{key_eq}, and substituting
$t=t_0(1+x)$, we obtain for $x\to 0$ that
\begin{align}
\frac{f\(\sqrt{1+\dot{z}^2}-\sqrt{f+\dot{z}^2}\)}{\sqrt{f+\dot{z}^2}+\dot{z}}\frac{d-1}{2z_0}\frac{B_0}{\(-x+\frac{\dot{z}}{\sqrt{f+\dot{z}^2}}x\pm\eps\)^c}=\frac{2B_1(c-1)}{\(-x-\frac{\dot{z}}{\sqrt{f+\dot{z}^2}}x\pm\eps\)^c},
\end{align}
which gives
\begin{align}
B_1=B_0\frac{d-1}{4z_0(c-1)}\frac{f\(\sqrt{1+\dot{z}^2}-\sqrt{f+\dot{z}^2}\)}{\sqrt{f+\dot{z}^2}+\dot{z}}\(\frac{\sqrt{f+\dot{z}^2}+\dot{z}}{\sqrt{f+\dot{z}^2}-\dot{z}}\)^c
\, .
\end{align}

The second step of matching is to be done near the boundary point
$t=\tb_1$, which is the overlap region of the segments $(1,+)$ and $(1,-)$. The form of the most singular part of $\phi_{1,-}$ is determined such that $\phi=\phi_{1,+}+\phi_{1,-}$ is free of source, i.e. $\phi$ contains only the vev term $z^d$ there.
It is useful to separate the positive and negative contributions and look at them in frequency representation. The near-boundary solution of $\phi_\pm$ are given by
\begin{align}
\text{For}\;\omg>0:  \no
&\phi_-\sim z^{\frac{d}{2}}H^{(1)}_{\frac{d}{2}}(\omg z) \, , \qquad
\phi_+\sim z^{\frac{d}{2}}H^{(2)}_{\frac{d}{2}}(\omg z), \\
\text{For}\;\omg<0 : \no
&\phi_-\sim z^{\frac{d}{2}}H^{(2)}_{\frac{d}{2}}(-\omg z) \, , \qquad 
\phi_+\sim z^{\frac{d}{2}}H^{(1)}_{\frac{d}{2}}(-\omg z).
\end{align}
In order to cancel the $z^0$ term near the boundary for the positive
frequency contribution, we need to have a $H^{(1)}_{\frac{d}{2}}(\omg
z)$ in $\phi_{1,-}$ for each $H^{(1)}_{\frac{d}{2}}(\omg z)$ in
$\phi_{1,+}$. On the other hand, the coordinate representation
$\phi_{1,+}^>$ and $\phi_{1,-}^>$ take the form 
\begin{gather}
\phi_{1,+}^>=\frac{z^{\frac{d-1}{2}}B_1}{\(-t+t'+i\eps-\Z\)^{c-1}}
\,, \quad \phi_{1,-}^>=\frac{z^{\frac{d-1}{2}}\#}{\(-t+t'+i\eps+\Z\)^{c-1}}\, ,
\end{gather}
with $\#$ to be determined by the matching. Note that $\Z\to z$ as
$z\to 0$. Using the asymptotic expansion of the Hankel function, we conclude that
\begin{align}
\phi_{1,-}^>=\frac{z^{\frac{d-1}{2}}e^{-i\pi c}B_1}{\(-t+t'+i\eps+\Z\)^{c-1}}.
\end{align}
A similar analysis of the negative frequency contribution leads to
\begin{align}
\phi_{1,-}^<=\frac{z^{\frac{d-1}{2}}e^{i\pi c}B_1}{\(-t+t'-i\eps+\Z\)^{c-1}}.
\end{align}
Repeating this process, we obtain  for the most singular part of the correlator
\begin{align}\label{cn_condition}
&\phi_{n,+}^>=\frac{z^{\frac{d-1}{2}}e^{-i\pi c(n-1)}B_n}{\(-t+\tb_n+i\eps-\Z\)^{c-n}},\quad \phi_{n,+}^<=-\frac{z^{\frac{d-1}{2}}e^{i\pi c(n-1)}B_n}{\(-t+\tb_n-i\eps-\Z\)^{c-n}}, \no
&\phi_{n,-}^>=\frac{z^{\frac{d-1}{2}}e^{-i\pi cn}B_n}{\(-t+\tb_n+i\eps+\Z\)^{c-n}},\quad \phi_{n,-}^<=-\frac{z^{\frac{d-1}{2}}e^{i\pi cn}B_n}{\(-t+\tb_n-i\eps+\Z\)^{c-n}}.
\end{align}
$B_n$ is to be determined from the recursion relation 
\begin{align}\label{B_recursion}
B_n=B_{n-1}\frac{d-1}{4z(c-n)}\frac{f\(\sqrt{1+\dot{z}^2}-\sqrt{f+\dot{z}^2}\)}{\sqrt{f+\dot{z}^2}+\dot{z}}\(\frac{\sqrt{f+\dot{z}^2}+\dot{z}}{\sqrt{f+\dot{z}^2}-\dot{z}}\)^{c-n+1},
\end{align}
where all the quantities are evaluated at $t=t_{n-1},\,z=z_{n-1}$. 

Now that we have obtained the most singular part of $\phi_{n,\pm}$ close to the segments $(n,\pm)$, the next step to reconstruct the most singular part of the boundary retarded two-point correlator in the vicinity of $\tb_n$. This is done as follows: We first write $\phi_{n,\pm}^{>,<}$ in frequency representation,
\begin{align}\label{freq_rep}
&\omg>0: \no
&\phi_{n,+}^>=\int_0^\infty d\omg
g_{n,+}^>(\omg)z^{\frac{d}{2}}H^{(2)}_{\frac{d}{2}}(\omg z)e^{-i\omg
  t} \, , \no
&\phi_{n,-}^>=\int_0^\infty d\omg g_{n,-}^>(\omg)z^{\frac{d}{2}}H^{(1)}_{\frac{d}{2}}(\omg z)e^{-i\omg t}, \\
&\omg<0: \no
&\phi_{n,+}^<=\int_{-\infty}^0d\omg
g_{n,+}^<(\omg)z^{\frac{d}{2}}H^{(1)}_{\frac{d}{2}}(-\omg z)e^{-i\omg
  t} \, , \no
&\phi_{n,-}^<=\int_{-\infty}^0d\omg g_{n,-}^<(\omg)z^{\frac{d}{2}}H^{(2)}_{\frac{d}{2}}(-\omg z)e^{-i\omg t}.
\end{align}
In the UV limit, we can use asymptotic expansion of the Hankel functions on the right hand sides of \eqref{freq_rep}, which become Laplace transforms (for negative frequency contribution, a change of variable is needed). These are easily inverted to give
\begin{align}\label{weight}
&g_{n,+}^>(\omg)=g_{n,-}^>(\omg)=\sqrt{\frac{\pi\omg}{2}}e^{i\omg
  \tb_n}e^{-i\pi cn}e^{\frac{i\pi
    n}{2}}B_n\frac{\omg^{c-n-1}}{\Gamma(c-n)} \, ,\no
&g_{n,+}^<(\omg)=g_{n,-}^<(\omg)=-\sqrt{\frac{-\pi\omg}{2}}e^{i\omg
  \tb_n}e^{i\pi cn}e^{-\frac{i\pi
    n}{2}}B_n\frac{(-\omg)^{c-n-1}}{\Gamma(c-n)} , 
\end{align}
with $c$ given by \eqref{c}. 
Finally, the most singular part of the two-point correlator is
obtained by taking the coefficients of the $z^d$ term and convoluting
with the weight in \eqref{weight},
\begin{align}
G_{n,+}^>&=\int_0^\infty d\omg \, \frac{e^{\frac{i\pi
      d}{2}}}{i\sin\frac{\pi
    d}{2}}\frac{1}{\Gamma(\frac{d}{2}+1)}\(\frac{\omg}{2}\)^{\frac{d}{2}}e^{-i\omg
  t}\, g_{n,+}^>(\omg) \, , \no
G_{n,-}^>&=\int_0^\infty d\omg \, \frac{-e^{\frac{-i\pi
      d}{2}}}{i\sin\frac{\pi
    d}{2}}\frac{1}{\Gamma(\frac{d}{2}+1)}\(\frac{\omg}{2}\)^{\frac{d}{2}}e^{-i\omg
  t}\, g_{n,-}^>(\omg)  \, , \\
G_{n,+}^<&=\int_{-\infty}^0 d\omg \, \frac{-e^{\frac{-i\pi
      d}{2}}}{i\sin\frac{\pi
    d}{2}}\frac{1}{\Gamma(\frac{d}{2}+1)}\(\frac{-\omg}{2}\)^{\frac{d}{2}}e^{-i\omg
  t}\, g_{n,+}^<(\omg) \, , \no
G_{n,-}^<&=\int_{-\infty}^0 d\omg \, \frac{e^{\frac{i\pi
      d}{2}}}{i\sin\frac{\pi
    d}{2}}\frac{1}{\Gamma(\frac{d}{2}+1)}\(\frac{-\omg}{2}\)^{\frac{d}{2}}e^{-i\omg
  t}\, g_{n,-}^<(\omg) \, .
\end{align}
It is easy to perform the integral and we combine the results from positive and negative frequency contribution as follows,
\begin{align}\label{Gn}
&G_n^>=G_{n,+}^>+G_{n,-}^>=\frac{\sqrt{\pi}}{2^{c-1}\Gm(\frac{d}{2}+1)}\frac{\Gm(2c-n)}{\Gm(c-n)}\frac{B_ne^{-i\pi c(n-1)}}{(-t+\tb_n+i\eps)^{2c-n}} \no
&G_n^<=G_{n,+}^<+G_{n,-}^<=-\frac{\sqrt{\pi}}{2^{c-1}\Gm(\frac{d}{2}+1)}\frac{\Gm(2c-n)}{\Gm(c-n)}\frac{B_ne^{i\pi c(n-1)}}{(-t+\tb_n+i\eps)^{2c-n}}.
\end{align}
Note that the singularities of the correlator become milder as $n$ increases. In other words, every bouncing of the light ray on the shell lower the power of the singularities by one. This is a consequence of the same effect on the bulk scalar close to the segments depicted in Fig.\ref{bouncing}. Furthermore, there is also a suppression on the numerical factor $B_n$ by each bouncing as dictated by \eqref{B_recursion}.

\subsection{An explicit example: The quasi-static state}\label{sec_quasi}

In this section, we will test our algorithm with an example where explicit analytic computation is possible. Since the procedure outlined in the previous section is valid for all shell trajectory, we choose to work in the case of static shell: $\dot{z}=0$. This can be viewed as a quasi-static state studied in \cite{shell}.

Since the shell is static, modes with different frequency modes decouple, which allows for a simple treatment in frequency representation. We start with the matching condition \eqref{mat_cond}. Setting $\dot{z}=0$, we have
\begin{align}\label{FT}
\left\{\begin{array}{l}
\phi_f=\phi\\
\sqrt{f}\del_z\phi_f=\del_z\phi
\end{array}
\right.\overset{\text{F.T.}}{\Longrightarrow}
\left\{\begin{array}{l}
\pht_f=\frac{1}{\sqrt{f}}\pht\\
\sqrt{f}\del_z\pht_f=\frac{1}{\sqrt{f}}\del_z\pht
\end{array}
\right..
\end{align}
The right hand side is the Fourier transform of the left hand side. We have used a tilde to indicate quantities in the frequency space. When $\dot{z}=0$, the mismatch between the time coordinates above and below the shell is simply: $dt_f=\frac{1}{\sqrt{f}}dt$. This also leads to a corresponding mismatch in frequencies and the additional factor $\frac{1}{\sqrt{f}}$ on the right hand side of \eqref{FT} \cite{shell}. We will take the frequency above the shell to be $\omg$, thus the corresponding frequency below the shell is $\frac{\omg}{\sqrt{f}}$.

In the UV limit, $\pht$ satisfies $\del_z\pht-\frac{i\omg}{\sqrt{f}}\pht=\frac{d-1}{2z}\pht$. Plugging this into \eqref{FT}, we obtain
\begin{align}\label{pht_mat}
f\del_z\pht_f-i\omg\pht_f=\frac{d-1}{2z}\sqrt{f}\pht_f.
\end{align}
We again split the wave above the shell into ingoing and outgoing components,
\begin{align}\label{pht_split}
&\pht_f=A\pht_-+B\pht_+ \, , \no
&\text{where}\;\pht_\mp=z^{\frac{d-1}{2}}e^{\pm i\omg\Z}.
\end{align}
The ratio $\frac{A}{B}$ is readily determined from \eqref{pht_mat}. To
further simplify the calculation, we keep only terms to the leading
order in $f$ (or $1-z_s$, where $z_s$ is the radial position of the shell). We obtain
\begin{align}\label{AB}
\frac{A}{B}=\frac{-4i\omg}{\sqrt{f}(d-1)}\frac{\pht_+}{\pht_-}=\frac{-4i\omg}{\sqrt{f}(d-1)}e^{-2i\omg\Z}.
\end{align}
To calculate the two-point function, we need an expansion near
$z=0$. Note that $\pht_\pm$ defined in \eqref{pht_split} is not valid
as $z\to 0$. However, they may be matched to the near-boundary
solution which is given by Hankel functions,
\begin{align}
&\text{For}\; \omg>0: \no
&\pht_-\to z^{\frac{d}{2}}H^{(1)}_{\frac{d}{2}}(\omg z)e^{\frac{i\pi c}{2}}, \quad\quad
\pht_+\to z^{\frac{d}{2}}H^{(2)}_{\frac{d}{2}}(\omg z)e^{\frac{-i\pi c}{2}}, \\
&\text{For}\; \omg<0: \no
&\pht_-\to z^{\frac{d}{2}}H^{(2)}_{\frac{d}{2}}(-\omg z)e^{\frac{-i\pi c}{2}}, \quad\quad
\pht_+\to z^{\frac{d}{2}}H^{(1)}_{\frac{d}{2}}(-\omg z)e^{\frac{i\pi c}{2}}.
\end{align}
With \eqref{AB} and the mapping above, we may write down the retarded correlator for separate contributions from positive and negative frequencies as follows,
\begin{align}\label{Gr_omg}
&G_R^>=\frac{1}{2\pi}\int_0^\infty d\omg
e^{-i\omg(t-t')}\frac{-e^{-i\frac{\pi d}{2}}Ae^{\frac{i\pi
      c}{2}}+e^{i\frac{\pi d}{2}}Be^{\frac{-i\pi
      c}{2}}}{Ae^{\frac{i\pi c}{2}}-Be^{\frac{-i\pi
      c}{2}}}\frac{\Gamma(1-\frac{d}{2})}{\Gamma(1+\frac{d}{2})}\(\frac{\omg}{2}\)^d
\, , \\
&G_R^<=\frac{1}{2\pi}\int_{-\infty}^0 d\omg e^{-i\omg(t-t')}\frac{e^{i\frac{\pi d}{2}}Ae^{\frac{-i\pi c}{2}}-e^{-i\frac{\pi d}{2}}Be^{\frac{i\pi c}{2}}}{-Ae^{\frac{-i\pi c}{2}}+Be^{\frac{i\pi c}{2}}}\frac{\Gamma(1-\frac{d}{2})}{\Gamma(1+\frac{d}{2})}\(\frac{-\omg}{2}\)^d.
\end{align}
A direct integration of \eqref{Gr_omg} is not possible. Note that
$\frac{B}{A}$ is suppressed by $\frac{1}{\omg}$ in the UV limit. We
may perform a series expansion in $\frac{B}{A}$ of the integrand using
\begin{align}
&G_R^>=\frac{1}{2\pi}\int_0^\infty d\omg \,
e^{-i\omg(t-t')}\(-e^{-\frac{i\pi
    d}{2}}\)\bigg[1+\sum_{n\ge1}\(1-e^{i\pi d}\)\(\frac{B}{A}e^{-i\pi
  c}\)^n\bigg]\frac{\Gamma(1-\frac{d}{2})}{\Gamma(1+\frac{d}{2})}\(\frac{\omg}{2}\)^d
\, ,\\
&G_R^<=\frac{1}{2\pi}\int_{-\infty}^0 d\omg \, e^{-i\omg(t-t')}\(-e^{-\frac{i\pi d}{2}}\)\bigg[1+\sum_{n\ge1}\(1-e^{-i\pi d}\)\(\frac{B}{A}e^{i\pi c}\)^n\bigg]\frac{\Gamma(1-\frac{d}{2})}{\Gamma(1+\frac{d}{2})}\(\frac{-\omg}{2}\)^d.
\end{align}
The lowest order  terms, i.e.~the $1$ in the square brackets, simply give the vacuum retarded correlator. The $n$-th order terms precisely give us the most singular part of the retarded correlator,
\begin{align}\label{Gnr_omg}
&G_{n,R}^>=\frac{i\sin\frac{\pi
    d}{2}}{2^d\pi}\frac{\Gm(1-\frac{d}{2})}{\Gm(1+\frac{d}{2})}\(\frac{\sqrt{f}(d-1)}{4}\)^n\frac{\Gm(d-n+1)e^{-i\pi
    c(n-1)}}{\(-t+t'+2n\Z+i\eps\)^{d-n+1}} \, , \no
&G_{n,R}^<=-\frac{i\sin\frac{\pi d}{2}}{2^d\pi}\frac{\Gm(1-\frac{d}{2})}{\Gm(1+\frac{d}{2})}\(\frac{\sqrt{f}(d-1)}{4}\)^n\frac{\Gm(d-n+1)e^{i\pi c(n-1)}}{\(-t+t'+2n\Z-i\eps\)^{d-n+1}}.
\end{align}
We note that the dependence on the temperature is not visible for the WKB approximation we are working in.

Let us calculate the same quantities by applying our algorithm
outlined in the previous section. The only input required for
\eqref{Gn} is the $B_n$. Working to the lowest order in $f$, we obtain from \eqref{B_recursion}
\begin{align}
B_n=\frac{d-1}{4(c-n)}B_{n-1}\sqrt{f}.
\end{align}
The initial condition for the recursion equation is $B_0=\frac{i}{2^c\pi}\frac{\Gamma(c)\Gamma(\frac{1}{2})}{\Gamma(\frac{d}{2})}$.
Noting that $\tb_n=t'+2n\Z$ in case of a static shell, we can show that \eqref{Gnr_omg} and \eqref{Gn} are in perfect agreement!

\subsection{Thermalizing state}

Now that our algorithm has survived the non-trivial test, we can apply it to the more complicated case of gravitational collapse, with trajectory of the shell satisfying 
\begin{align}
\frac{dz}{dt_f}=\frac{2f\sqrt{\frac{1}{4}\(bz^4+\frac{1}{bz_h^4}\)^2-1}}{\frac{1}{bz_h^4}-bz^4}.
\end{align}
We may deduce the properties of the most singular part of the
correlator already from \eqref{Gn}. We will specialize to $d=4$ from
now on. For a given time $t'$, the spatial integrated retarded
correlator $\int d^3x\, \theta(t-t')[O(t'),O(t)]$ contains singularities as $t\to\tb_n$, with $\tb_n$ being the $n$-th boundary point on the trajectory of the bouncing light ray. The most singular part of the correlator takes the  generic form
\begin{align}
&G_n^>=\frac{A_ne^{-i\pi c(n-1)}}{(-t+\tb_n+i\eps)^{5-n}},\quad\quad G_n^<=-\frac{A_ne^{i\pi c(n-1)}}{(-t+\tb_n+i\eps)^{5-n}}.
\end{align}
The numerical factor $A_n$ has to be determined for a given
trajectory. The condition of the divergence of \eqref{cn_condition}
implies $n<c=\frac{5}{2}$, which is the reason we have mapped out the
number of bouncings only upto $n=2$. The $n=0$ case gives simply the
lightcone singularity. The nontrivial singularities appear for
$n\ge1$. We have shown in Section 3.1 that $\tb_n\to+\infty$ as $t'\to
T_n(z_s)$ from below, which means the retarded correlator will be free
of singularities when $t'>T_1(z_s)$. Therefore we propose to use
$T_1(z_s)$ as our definition of the thermalization time $t_\mathrm{th}$. Let us
discuss the connection to heavy ion physics. Restoring units, we have
\begin{align}
t_{\mathrm{th}}=\frac{T_1(\pi Tz_s)}{\pi T},
\end{align}
where $T$ corresponds to the initial temperature of the quark-gluon
plasma (QGP) formed in heavy ion collisions and $z_s$ is related to the saturation scale of the nucleus. We see that the thermalization time is inversely proportional to the initial temperature and the numerical factor $T_1$ involves an interplay between the temperature and saturation scale, which is ultimately determined by the collision energy and size of the nucleus.

In  Kruskal coordinates, the light ray will always have at least one
bouncing. But for large $t'$, the bouncing will occur only after the
shell crosses the horizon, thus the light ray will not be able to
return to the boundary. From the point of view of  Schwarzschild
coordinates, this corresponds to the case where the light ray never
reaches the falling shell. Therefore for sufficiently large $t'$, we expect the singularities of the correlator to be identical to the singularities of thermal correlator evaluated in the background of an eternal black hole. In the next section, we will find the singularities in thermal correlator actually appears in the complex $t$ plane.

\section{Singularities in thermal correlator}

\begin{figure}
\includegraphics[width=0.8\textwidth]{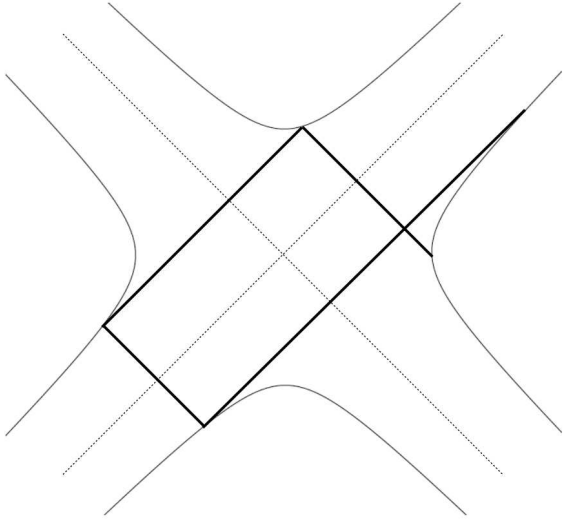}
\caption{\label{fig:kruskal} Bouncing of a null geodesic in the AdS black hole
in Kruskal coordinates. The geodesics starts on the AdS boundary on
the right and bounces off the future singularity, the second AdS
boundary and the past singularity back to the boundary. A similar set-up has first been discussed in \cite{hoyos}.}
\end{figure}

We now move to the second part of this paper where
we establish a relation between the singularities of the retarded correlator
in coordinate space and the quasi-normal models (QNM). We first do
this for a field theory in thermal equilibrium, which is dual to the
AdS-Schwarzschild background. We obtain explicit expressions for
thermal field correlator in coordinate space. These expressions allow
us to locate singularities in the complex time plane, which turns out
to be closely related to the QNM. The first indication of this
connection was discussed in \cite{hoyos}, where a geometric optics
picture has been used to derive the asymptotic QNM. It was assumed
that the coordinate space correlator has periodic singularities in
complex time, with a complex period determined by the time a light ray
needs to make a loop by bouncing in the full Penrose diagram, as shown
in figure \ref{fig:kruskal}.

Applying
this picture here, we may draw the following conclusion from our study of the gravitational collapse model: As we increase the initial time $t'$, the unequal-time correlator probes different stages of the thermalization process dual to the gravitational collapse model. The resulting singularities of the correlator show a shift from the real time to complex time. We will solidify this picture by explicit evaluations of the correlators in the BTZ black hole and AdS$_5$-Schwarzschild black hole. We will see that singularities are indeed present in both cases.

\subsection{Unequal time correlator from BTZ black hole}

We are interested in the singularities of correlator in coordinate space. The black hole background preserves time translational symmetry, therefore the unequal time correlator only depends on the difference of the insertion times of the operator, $G^R(t,t')=G^R(t-t')$. Below we will  set $t'=0$ for convenience. We start with the correlator in momentum representation,
\begin{align}\label{G>}
G^R(\omg)=\int dt \,  e^{i\omg t}G^R(t).
\end{align}
We will obtain $G^R(\omg)$ by solving massless scalar wave equation in
the BTZ background.
In the reduced unit $2\pi T=1$, the BTZ black hole takes the  form
\begin{align}\label{btz}
ds^2=\frac{-(1-z^2)dt^2+dz^2/(1-z^2)+d\phi^2}{z^2},
\end{align}
where $\phi$ is periodic $\phi\sim\phi+2\pi$. The solution of the
massless scalar satisfying the ingoing boundary condition at the
horizon and approaching unity at the BTZ boundary is expressed in
terms of a hypergeometric function,
\begin{align}\label{massless}
\Phi(z)=\frac{\Gm(1-\frac{i\omg}{2})^2}{\Gm(1-i\omg)}\(1-z^2\)^{-\frac{i\omg}{2}}z^2F\(1-\frac{i\omg}{2},1-\frac{i\omg}{2};1-i\omg;1-z^2\).
\end{align}
Note that we consider a solution without $\phi$-dependence, which
corresponds to the spatially integrated correlator where an
integration over $\phi$ is implicit. \eqref{massless} leads to the retarded correlator 
\begin{align}
G^R(\omg)=i\omg-\omg^2\psi(1-\frac{i\omg}{2}).
\end{align}

All the poles of $G_R(\omg)$ lie in the lower half plane at
$\omg=-i(2l+2)$, which means the correlator is nonvanishing only for
$\mathrm{Re} \, t \, > 0$. In the latter case, we close the contour clockwise and pick up the poles in the lower half plane. It is easy to sum over the residues. We obtain
\begin{align}\label{G>btz}
G^R(t)=-2\sum_{l=0}^\infty e^{-(2l+2)t}(2l+2)^2=\frac{2\cosh t}{\sinh^3t}.
\end{align}

From the picture of a light ray bouncing in a confining box of the
Penrose diagram as shown in figure \ref{fig:kruskal}, we expect
singularities to have a period of $\Delta t=i\pi$, which is the time a
light ray needs to finish a loop. Explicit expression for the
correlator \eqref{G>btz} indeed shows singularities at $t=i\pi n$ for
$n\in Z$. Strictly speaking, we need to give $t$ an infinitesimal real
part to justify $\mathrm{Re} \, t\, >0$. This will not be necessary
for the case of AdS$_5$-Schwarzschild black hole below. This
difference arising from the global structure of spacetime between BTZ space
and higher-dimensional AdS-Schwarzschild space has been discussed at length in \cite{hubeny}.

\subsection{Unequal time correlator from AdS$_5$ Schwarzschild black hole}

Now let us look at the correlator in the AdS$_5$ Schwarzschild background. The unequal-time correlator is again given by
\begin{align}\label{G>ads5}
G^R(t)=\frac{1}{2\pi}\int \! d\omg \, e^{-i\omg t} \, G^R(\omg).
\end{align}
The derivation of the relevant part of $G^R(\omg)$ can be found in the appendix.
Again it is nonvanishing for $\mathrm{Re} \, t \, >0$. We close the contour clockwise and pick up poles in the lower half plane. The poles from the QNM is expected to give singularity of $G^R(t)$ in the complex $t$ plane. Since the explicit expression for $G^R(\omg)$ is not available, we will try to evaluate \eqref{G>ads5} approximately. This amounts to summing over the contributions from poles with large $|\omg|$. The analytic expressions of asymptotic QNM for AdS$_5$ Schwarzschild have been obtained in \cite{NS,starinets}:
\begin{align}
\omg_n=\frac{n\pi+\theta}{x_0},\;\;\omg_n=\frac{n\pi+\theta}{\tilde{x}_0},
\end{align}
where $x_0=\frac{1+i}{4T}$, $\tilde{x}_0=\frac{-1+i}{4T}$ and
$\theta=\frac{5\pi}{4}+\frac{\ln 2i}{2i}$\footnote{Our definition of
  $x_0$ differs from that of \cite{NS} by complex conjugation}.
To proceed further, we need to know the residues at the QNM, all of
which are simple poles. This can be achieved by 
generalizing the work \cite{NS}. We collect necessary steps in the
appendix. We obtain for the residues of $G^R(\omg)$ 
\begin{gather}
\text{res}(\omg_n=\frac{n\pi+\theta}{x_0})=-\frac{\pi\omg_n^4}{32x_0}
\, , \qquad 
\text{res}(\omg_n=\frac{n\pi+\theta}{\tilde{x}_0})=-\frac{\pi\omg_n^4}{32\tilde{x}_0}.
\end{gather}
Combined with the fact that $G^R(\omg\gg T)\sim\omg^4\ln\omg$, we  make an educated guess for $G^R(\omg)$,
\begin{align}\label{Grapprox}
G^R(\omg)=-\frac{\omg^4}{32}\(\psi(\frac{\theta-\omg x_0}{\pi})+\psi(\frac{\theta-\omg \tilde{x}_0}{\pi})\).
\end{align}
This approximate expression is also consistent with the symmetry
$G_R(\omg)=G_R^*(-\omg^*)$, which is present for $|\omg|\gg 1$, when we may ignore the dependence on the offset $\theta$.
Now we are in a position to evaluate \eqref{G>ads5} with the approximate expression \eqref{Grapprox}. 

The contribution to $G^R(t)$ from the QNM at
$\omg_n=\frac{n\pi+\theta}{x_0}$ is calculated to be
\begin{align}\label{qnm_pole}
\frac{i}{32}\sum_ne^{-i\omg_n t}\omg_n^4\frac{\pi}{x_0}&=\frac{i}{32}e^{-\frac{i(n\pi+\theta)t}{x_0}}\(\frac{n\pi+\theta}{x_0}\)^4\frac{\pi}{x_0} \no
&=\frac{i}{32}e^{-\frac{i\theta
    t}{x_0}}\(\frac{\pi}{x_0}\)^5\Phi(e^{-\frac{i\pi t}{x_0}},-
4,\frac{\theta}{\pi})\, \deq \, \frac{4!i}{32}e^{2m_1\theta i}\frac{1}{(it+2m_1x_0i)^5},
\end{align}
where $\Phi$ is the Lerch transcendental function and $m_1\in Z$. The
symbol $\deq$ means that the most singular parts of the two sides are
equal. The contribution from the QNM at
$\tilde\omg_n=\frac{n\pi+\theta}{\tilde{x}_0}$ to $G^R(t)$ can be
easily obtained by the substitution $x_0\to\tilde{x}_0$ in
\eqref{qnm_pole}, with the  singular part
\begin{align}\label{qnm2_pole}
\frac{i}{32}\sum_ne^{-i\tilde\omg_n
  t}\tilde\omg_n^4\frac{\pi}{\tilde{x}_0}\, \deq \, \frac{4!i}{32}e^{2m_2\theta i}\frac{1}{(it+2m_2\tilde{x}_0i)^5},
\end{align}
where $m_2\in Z$. In the region $\mathrm{Re} \, t\, >0$, we find
singularities at $t=-2m_1x_0$ and at $t=-2m_2\tilde{x}_0$ for $m_1<0,\,m_2>0$. The singularities indeed have periods dictated by the dual periods of the QNM $\frac{\pi}{x_0}$ and $\frac{\pi}{\tilde{x}_0}$. This completes our discussion of the singularities in thermal field correlator and confirms the existence of singularities in the complex time plane.

\section{Singularities in unequal-time correlator from evolution of QNM}\label{sec_qnm}

The explicit calculation in the previous section has taught  us an
important lesson: The singularities of the correlator in the complex
time plane are closely related to the structure of the quasi-normal
modes (QNM). In this
section, we aim at improving our understanding of the singularities of
the unequal time correlator by studying the evolution of QNM in the
gravitational collapse process. This is only possible for the
quasi-static state, which is defined by a sequence of states
corresponding to the shell held at different positions above the
horizon. As the shell is lowered to the horizon, we expect to observe
features of the thermal field. While the sequence of states does not
correspond to physical state undergoing thermalization, it does help
us to gain more insight to the thermalization process. We specialize
to $d=2$, corresponding to the formation of BTZ black hole, which we
have a better analytic control on. We study the evolution of the
singularities in the unequal time correlator from the dual evolution
of the QNM. We reproduce the singularities of \eqref{Gnr_omg} in real
$t$ from the normal modes, which are a subset of the QNM. We  also explore the possibility of having singularities in complex $t$ from the other QNM.

For the quasi-static state, we can look at the retarded correlator
$G^q_R$ in momentum space, with the superscript $q$ indicating it
corresponds to a quasi-static state. We will study its QNM in the
whole complex plane. We need to work out the matching condition on the
shell in momentum space, which is valid for arbitrary complex
$\omg$. The starting point is again the RHS of \eqref{FT},
\begin{align}\label{matching_complex}
\left\{\begin{array}{l}
\pht_f=\frac{1}{\sqrt{f}}\pht\\
\sqrt{f}\del_z\pht_f=\frac{1}{\sqrt{f}}\del_z\pht
\end{array}
\right.,
\end{align}
with $\omg$ and $\frac{\omg}{\sqrt{f}}$ being the frequency above and below the shell respectively. For complex $\omg$, the scalar below the shell is given by
\begin{align}\label{besselk}
\pht= zK_1(e^{-\frac{i\pi}{2}}\frac{\omg}{\sqrt{f}} z).
\end{align}
We may check that for $\omg>0$, this reduces to the real expression
$zH_1^{(1)}(\frac{\omg}{\sqrt{f}} z)$. Conversely, for $\omg<0$ we find
$zH_1^{(2)}(-\frac{\omg}{\sqrt{f}} z)$. There is a branch cut on the negative imaginary axis, which is crucial in reproducing the retarded correlator in coordinate space.

The scalar above the shell is a linear combination of ingoing and
outgoing waves $\pht_f=A\pht_-+B\pht_+$. When shell is sufficient
close to the horizon, $\pht_\mp$ is approximated by a series solution near the horizon. We choose
\begin{align}\label{near_horizon}
\pht_\mp=(1-z)^{\frac{\mp i\omg}{2}}e^{\pm i\omg\frac{\ln2}{2}}+O(1-z).
\end{align}
The $z$-independent factor is introduced such that $\pht_\mp$ reduces
to \eqref{pht_split} as $z\to1$. Now inserting this into
\eqref{matching_complex} and performing a series expansion in $f$ or
effectively in $1-z$, we find that the corrections to series solution
$\pht_\mp$ are subleading and an asymptotic expansion of \eqref{besselk} gives
\begin{align}\label{A_B}
&A=e^{\frac{i\omg}{\sqrt{f}}}f^{-1+\frac{i\omg}{2}}2^{\frac{1}{2}}\(e^{-\frac{i\pi}{2}}\omg\)^{\frac{1}{2}}\sqrt{\pi}f^{\frac{3}{4}}
\, ,\no
&B=e^{\frac{i\omg}{\sqrt{f}}}f^{-1-\frac{i\omg}{2}}\frac{2^{-\frac{3}{2}}\(e^{-\frac{i\pi}{2}}\omg\)^{\frac{1}{2}}\sqrt{\pi}f^{\frac{5}{4}}i}{\omg}.
\end{align}
We note that although the branch cut along the negative imaginary axis
exists in both $A$ and $B$, it is absent in their ratio
\begin{align}\label{ratio_complex}
\frac{B}{A}=f^{-i\omg}\frac{i\sqrt{f}}{4\omg}.
\end{align}
The ratio agrees with what we obtained for real $\omg$ in \eqref{AB}
in the limit $f\to 0$. A more general expression can be obtained for
an arbitrary choice of the ingoing/outgoing wave solutions $\pht_-$/$\pht_+$ in $d$ dimensions,
\begin{align}\label{ratio_general}
\frac{B}{A}=\frac{i(d-1)\sqrt{f}}{4\omg}\frac{\pht_-}{\pht_+},
\end{align}
to be evaluated at a position of the shell sufficient close to the
horizon. W only need \eqref{ratio_complex} for our purpose, from which
we may derive the corresponding retarded correlator for a quasi-static state,
\begin{align}\label{Gr_complex}
G_R^q(\omg)=\frac{\pht_f^d}{\pht_f^0}=\frac{\pht_-^d+\frac{B}{A}\pht_+^d}{\pht_-^0+\frac{B}{A}\pht_+^0},
\end{align}
where the superscripts $0$ and $d$ denote the source and vev of the corresponding quantities. By construction, we have $\pht_-^d=G_R(\omg)\pht_-^0$ and $\pht_+^d=G_A(\omg)\pht_+^0$ up to contact terms. Therefore, we obtain
\begin{align}\label{Gr_complex2}
G_R^q(\omg)=\frac{\pht_-^0G_R(\omg)+\frac{B}{A}\pht_+^0G_A(\omg)}{\pht_-^0+\frac{B}{A}\pht_+^0}.
\end{align}

Now we use exact ingoing/outgoing wave solutions in BTZ black hole background,
\begin{align}\label{exact}
\pht_\mp=\(1-z^2\)^{\frac{\mp i\omg}{2}}z^2F\(1\mp\frac{i\omg}{2},1\mp\frac{i\omg}{2};1\mp i\omg;1-z^2\).
\end{align}
The normalizations in \eqref{exact} are chosen to match the near-horizon behavior in \eqref{near_horizon}. It is not difficult to obtain from \eqref{exact} that
\begin{align}\label{phi0}
&\pht_-^0=\frac{\Gm\(1-i\omg\)}{\Gm\(1-\frac{i\omg}{2}\)^2},\quad\quad \pht_+^0=\frac{\Gm\(1+i\omg\)}{\Gm\(1+\frac{i\omg}{2}\)^2},\\
&G_R(\omg)=i\omg-\omg^2\psi\(1-\frac{i\omg}{2}\),\quad G_A(\omg)=-i\omg-\omg^2\psi\(1+\frac{i\omg}{2}\),
\end{align}
where we have suppressed the overall numerical factor in $G_R$ and
$G_A$. Inserting \eqref{phi0} and \eqref{ratio_complex} into \eqref{Gr_complex2}, we obtain
\begin{align}\label{Gr_exp}
G_R^q(\omg)=\frac{\frac{\Gm\(1-i\omg\)}{\Gm\(1-\frac{i\omg}{2}\)^2}G_R(\omg)-f^{-i\omg}\frac{\sqrt{f}}{4i\omg}\frac{\Gm\(1+i\omg\)}{\Gm\(1+\frac{i\omg}{2}\)^2}G_A(\omg)}{\frac{\Gm\(1-i\omg\)}{\Gm\(1-\frac{i\omg}{2}\)^2}-f^{-i\omg}\frac{\sqrt{f}}{4i\omg}\frac{\Gm\(1+i\omg\)}{\Gm\(1+\frac{i\omg}{2}\)^2}}.
\end{align}
Now we are in the position to investigate the structure of QNM
corresponding to the retarded correlator \eqref{Gr_exp}. First we note
that possible QNM from $G_R(\omg)$ and $G_A(\omg)$ at $\omg=2i(n+1)$
and $\omg=-2i(n+1)$ do not arise since  the prefactors
$\frac{\Gm\(1-i\omg\)}{\Gm\(1-\frac{i\omg}{2}\)^2}$ and
$\frac{\Gm\(1+i\omg\)}{\Gm\(1+\frac{i\omg}{2}\)^2}$ evaluate to zero,
giving rise to a finite result. Therefore, we conclude that the only
QNM arise when the the denominator vanishes.  It is useful to write \eqref{Gr_exp} as
\begin{align}
G_R^q(\omg)=\frac{\frac{\Gm\(1+\frac{i\omg}{2}\)^2\Gm\(1-i\omg\)}{\Gm\(1-\frac{i\omg}{2}\)^2\Gm\(1+i\omg\)}G_R(\omg)-f^{-i\omg}\frac{\sqrt{f}}{4i\omg}G_A(\omg)}{\frac{\Gm\(1+\frac{i\omg}{2}\)^2\Gm\(1-i\omg\)}{\Gm\(1-\frac{i\omg}{2}\)^2\Gm\(1+i\omg\)}-f^{-i\omg}\frac{\sqrt{f}}{4i\omg}}.
\end{align}
Let us consider asymptotic QNM with $|\omg|\gg 1$. The limit $\lim_{\omg\to\infty}\frac{\Gm\(1+\frac{i\omg}{2}\)^2\Gm\(1-i\omg\)}{\Gm\(1-\frac{i\omg}{2}\)^2\Gm\(1+i\omg\)}$ depends on the argument of $\omg$. The role of the dependence of the correlator on the argument of $\omg$ has been emphasized in \cite{festuccia}. We first look at the limit along the real axis,
\begin{align}\label{asymp_real}
\frac{\Gm\(1+\frac{i\omg}{2}\)^2\Gm\(1-i\omg\)}{\Gm\(1-\frac{i\omg}{2}\)^2\Gm\(1+i\omg\)}=
\left\{\begin{array}{ll}
i4^{-i\omg}& \omg\to+\infty\\
-i4^{-i\omg}& \omg\to-\infty
\end{array}
\right..
\end{align}
Using \eqref{asymp_real}, we  find the asymptotic roots $\omg=a+ib$ with $|\omg|\gg 1$ for fixed $f$,
\begin{align}\label{root_real}
&a\approx-\frac{(2n-1)\pi}{\ln\frac{f}{4}},\; \quad
b\approx\frac{\ln\frac{4a}{f}}{\ln\frac{f}{4}} \, , \no
&a\approx\frac{2n\pi}{\ln\frac{f}{4}},\; \quad  b\approx\frac{\ln\frac{-4a}{f}}{\ln\frac{f}{4}},
\end{align}
with $n$ large positive integers. We find in both cases
$\frac{b}{a}\to 0$ as $n\to+\infty$ for fixed $f$, thus the roots are
asymptotically real, which justifies the use of \eqref{asymp_real}. We
identify the first and second lines of \eqref{root_real} as positive
and negative normal modes. These normal modes are equidistant with a
spacing $\Delta\omg=-\frac{2\pi}{\ln\frac{f}{4}}$, which gives rise to
the period $\Delta t=-\ln\frac{f}{4}$ in the singularities of the
unequal time correlator in coordinate space. Note that the period is
robust in that it is only governed by the phase factor
$\(\frac{f}{4}\)^{-i\omg}$. On the other hand, the period according to \eqref{Gnr_omg} is given by $2\Z\to-\ln(1-z)+\ln2\to -\ln\frac{f}{4}$ as $z\to1$. We see that the two approaches give consistent results.

\begin{figure}
\includegraphics[width=0.4\textwidth]{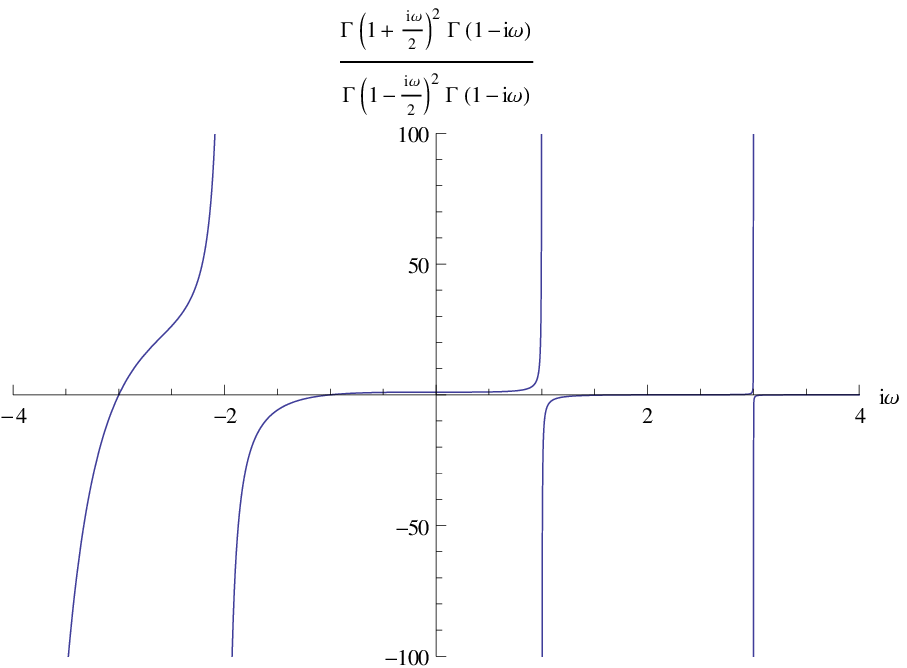}
\includegraphics[width=0.4\textwidth]{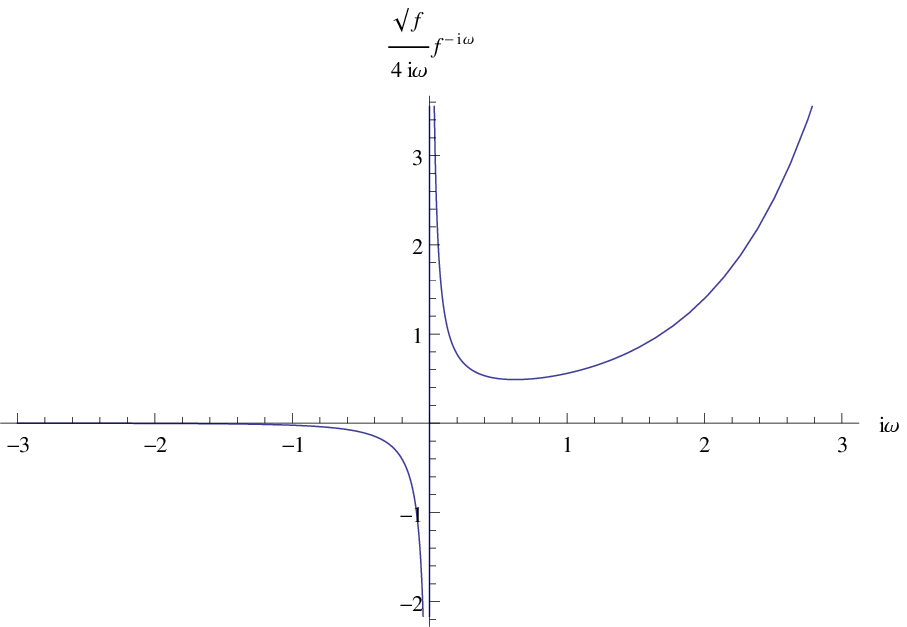}
\caption{\label{Fomg} The left plot shows $\frac{\Gm\(1+\frac{i\omg}{2}\)^2\Gm\(1-i\omg\)}{\Gm\(1-\frac{i\omg}{2}\)^2\Gm\(1+i\omg\)}$ versus $i\omg$ and the right plot shows $f^{-i\omg}\frac{\sqrt{f}}{4i\omg}$ versus $i\omg$.}
\end{figure}

Next we look for QNM along the imaginary axis. For purely imaginary $\omg$, the denominator is guaranteed to be real. It is convenient to study its behavior by plotting $\frac{\Gm\(1+\frac{i\omg}{2}\)^2\Gm\(1-i\omg\)}{\Gm\(1-\frac{i\omg}{2}\)^2\Gm\(1+i\omg\)}$ and $f^{-i\omg}\frac{\sqrt{f}}{4i\omg}$ as functions of $i\omg$. The former has infinities at $i\omg=2n-1$ and $i\omg=-2n$ and zeros at $i\omg=-(2n-1)$ and $i\omg=2n$ with $n$ positive integers. The latter has the simple asymptotics
\begin{align}
f^{-i\omg}\frac{\sqrt{f}}{4i\omg}\to
\left\{\begin{array}{ll}
0& i\omg\to-\infty\\
+\infty& i\omg\to+\infty
\end{array}
\right..
\end{align}
Fig.~\ref{Fomg} shows them in separate plots. 
It is easy to see that the QNM are given by their crossing points at
approximately $i\omg=-(2n-1)$ and $i\omg=2n-1$. The appearance of QNM
in the upper half plane might seem worrisome at the first sight. In
fact, this is just a reflection of the fact that our quasi-static
state is unstable against evolution toward equilibrium: From the bulk
point of view, we need an external force to keep the shell levitating
above the horizon. Note that for the thermal field, 
the QNM in the retarded and advanced correlators are at $i\omg=2n$ and
$i\omg=-2n$. Our QNM of quasi-static state agree with the thermal QNM
in the period. However, they differ in the absolute value by an
offset. Note that we are studying the QNM in the retarded correlator
for the quasi-static state. Nevertheless, we have found the QNM
corresponding to the retarded correlator at approximately
$i\omg=-(2n-1)$ and $i\omg=2n-1$, which are democratically distributed
along the positive and negative imaginary axis, i.e. there is no sign
of the retarded nature of the correlator. This is only visible when we
study the residues at the QNM. Since all the QNM are first order
poles, the corresponding residues are obtained by taking the derivative of the denominator with respect to $\omg$. The residues of $G^q_R(\omg)$ are given by 
\begin{align}\label{res}
i \,
\text{res}&=\frac{\frac{\Gm\(1+\frac{i\omg}{2}\)^2\Gm\(1-i\omg\)}{\Gm\(1-\frac{i\omg}{2}\)^2\Gm\(1+i\omg\)}G_R(\omg)-f^{-i\omg}\frac{\sqrt{f}}{4i\omg}G_A(\omg)}{\frac{d}{d(i\omg)}\bigg[\frac{\Gm\(1+\frac{i\omg}{2}\)^2\Gm\(1-i\omg\)}{\Gm\(1-\frac{i\omg}{2}\)^2\Gm\(1+i\omg\)}-f^{-i\omg}\frac{\sqrt{f}}{4i\omg}\bigg]}
\, , \no
&=\frac{f^{-i\omg}\frac{\sqrt{f}}{4i\omg}\(G_R(\omg)-G_A(\omg)\)}{\frac{d}{d(i\omg)}\bigg[\frac{\Gm\(1+\frac{i\omg}{2}\)^2\Gm\(1-i\omg\)}{\Gm\(1-\frac{i\omg}{2}\)^2\Gm\(1+i\omg\)}-f^{-i\omg}\frac{\sqrt{f}}{4i\omg}\bigg]},
\end{align}
We have inserted an $i$ to the residue for convenience. The residues
are found by first obtaining the QNM by solving 
\begin{gather}
\frac{\Gm\(1+\frac{i\omg}{2}\)^2\Gm\(1-i\omg\)}{\Gm\(1-\frac{i\omg}{2}\)^2\Gm\(1+i\omg\)}=f^{-i\omg}
\frac{\sqrt{f}}{4i\omg}\end{gather} and inserting them into
\eqref{res}. Table \ref{tab_res} shows the residues for several QNM, which encode the retarded nature of the correlator: The residue at $i\omg\approx-(2n-1)$ is always smaller than that at $i\omg\approx(2n-1)$ and their ratio decreases further as $f\to 0$.
\begin{table}
\begin{tabular}{cccc}
f=0.9& n=2& n=3& n=4\\
\hline
$i\, \text{res}(i\omg\approx-(2n-1))$& $-2.0\times10^{-5}$& $-4.6\times10^{-8}$& $-1.1\times10^{-10}$\\
$i\,  \text{res}(i\omg\approx2n-1)$& $0.72$& $9.3\times10^{-3}$& $8.4\times10^{-5}$\\
\hline  
ratio& $-2.8\times10^{-5}$& $-4.9\times10^{-6}$& $-1.3\times10^{-6}$\\
\hline\hline
f=0.1& n=2& n=3& n=4\\
\hline
$i \, \text{res}(i\omg\approx-(2n-1))$& $-4.2\times10^{-12}$& $-1.5\times10^{-18}$& $-5.4\times10^{-25}$\\
$i \, \text{res}(i\omg\approx2n-1)$& $8.8\times10^{-6}$& $2.3\times10^{-11}$& $3.3\times10^{-17}$\\
\hline  
ratio& $-4.8\times10^{-7}$& $-6.2\times10^{-8}$& $-1.6\times10^{-8}$\\
\end{tabular}
\caption{\label{tab_res} The residues of $G^q_R(\omg)$ at $i\omg\approx\pm(2n-1)$ for several $n$ at $f=0.9$ and $f=0.1$.}
\end{table}

With the above analysis, we
have shown the existence of QNM along the imaginary axis at symmetric
locations, but with asymmetric residues. Moreover,
we expect by inspecting the properties of QNM along the real axis that
there are also singularities in the complex $t$ plane with period
$\Delta t=i\pi$. However, it appears that in the WKB approach used
here, the possible singularities in the complex time plane are missed
out. Note that
the WKB approximation in Section.\ref{sec_quasi} is valid for real $\omg$ only. 

To see if singularities in the complex $t$ plane indeed exist, we need
to work a little harder to obtain the residues for the QNM. Analytic
calculation is possible asymptotically for large QNM with $n\gg 1$, due to the fact that the QNM asymptote to $i\omg=\pm(2n-1)$. At large $n$, the locations of the QNM are 
\begin{align}\label{qnm_shell}
&i\omg=-(2n-1)-\frac{f^{2n-\frac{1}{2}}}{4(2n-1)}\frac{\Gm(\frac{1}{2}+n)^2}{\Gm(\frac{3}{2}-n)^2\Gm(2n)(2n-2)!}
\, ,\no
&i\omg=(2n-1)-4(2n-1)f^{2n-\frac{3}{2}}\frac{\Gm(\frac{1}{2}+n)^2}{\Gm(\frac{3}{2}-n)^2\Gm(2n)(2n-2)!}.
\end{align}
Note the last term in both expressions in \eqref{qnm_shell} are corrections, which tend to zero rapidly as $n$ grows. Working to the lowest order in the corrections, we obtain the following results from \eqref{res},
\begin{align}\label{key}
&\text{For}\;i\omg\approx -(2n-1): \no
&\frac{f^{-i\omg}\frac{\sqrt{f}}{4i\omg}}{\frac{d}{d(i\omg)}\bigg[\frac{\Gm\(1+\frac{i\omg}{2}\)^2\Gm\(1-i\omg\)}{\Gm\(1-\frac{i\omg}{2}\)^2\Gm\(1+i\omg\)}-f^{-i\omg}\frac{\sqrt{f}}{4i\omg}\bigg]} 
=-\frac{\Gm(\frac{1}{2}+n)^2}{4f^{\frac{1}{2}-2n}(2n-1)\Gm(\frac{3}{2}-n)^2\Gm(2n)(2n-2)!}
\, , \\
&\text{For}\;i\omg\approx 2n-1: \no
&\frac{f^{-i\omg}\frac{\sqrt{f}}{4i\omg}}{\frac{d}{d(i\omg)}\bigg[\frac{\Gm\(1+\frac{i\omg}{2}\)^2\Gm\(1-i\omg\)}{\Gm\(1-\frac{i\omg}{2}\)^2\Gm\(1+i\omg\)}-f^{-i\omg}\frac{\sqrt{f}}{4i\omg}\bigg]} 
=4(2n-1)\frac{\Gm(\frac{1}{2}+n)^2}{f^{\frac{3}{2}-2n}\Gm(\frac{3}{2}-n)^2\Gm(2n)(2n-2)!}.
\end{align}
The remaining factor $G_R(\omg)-G_A(\omg)$ in \eqref{res} is also linear in the corrections, with the following expressions:
\begin{align}\label{Gra_shell}
G_R(\omg)-G_A(\omg)=\(i\omg\pm(2n-1)\)\(-\frac{\pi^2}{2}\)(2n-1)^2.
\end{align}
Inserting \eqref{key}, \eqref{Gra_shell} and \eqref{qnm_shell} into \eqref{res}, we obtain the residues 
\begin{align}\label{res_exp}
&\text{For}\;i\omg\approx -(2n-1): \no
&i\text{res}=-\frac{1}{32}\(\frac{\pi f^{2n-\frac{1}{2}}\Gm\(\frac{1}{2}+n\)^2}{\Gm\(\frac{3}{2}-n\)^2\Gm(2n)(2n-2)!}\)^2, \\
&\text{For}\;i\omg\approx 2n-1: \no
&i\text{res}=8(2n-1)^4\(\frac{\pi f^{2n-\frac{3}{2}}\Gm\(\frac{1}{2}+n\)^2}{\Gm\(\frac{3}{2}-n\)^2\Gm(2n)(2n-2)!}\)^2.
\end{align}
We may approximate the factor involving gamma functions as follows,
\begin{align}
\lim_{n\to+\infty}\frac{\Gm\(\frac{1}{2}+n\)^2}{\Gm\(\frac{3}{2}-n\)^2\Gm(2n)(2n-2)!}\approx\frac{8}{\pi\cdot 2^{4n}}.
\end{align}
For $\mathrm{Re}\, t\, <0$, we obtain the contribution from the QNM on the positive imaginary axis,
\begin{align}\label{ret<0}
G^q_R(t)\approx\sum_ne^{-i\omg t}i\text{res}(i\omg\approx-(2n-1))=\frac{e^{t}f^3}{2^7\(e^{2t+4\ln\frac{f}{4}}-1\)}.
\end{align}
For $\mathrm{Re} \, t\, >0$, the contribution from the QNM on the
negative imaginary axis has an additional enhancement factor
$(2n-1)^4$, and we end up with
\begin{align}\label{ret>0}
G^q_R(t)\approx\sum_ne^{-i\omg t}i\text{res}(i\omg\approx(2n-1))&=\frac{2^{13}e^{t}}{f^3}\Phi(e^{-2t}\(\frac{f}{4}\)^4,-4,\frac{1}{2}).
\end{align}
With our final expressions \eqref{ret<0} and \eqref{ret>0}, we may
answer the question about the singularities at complex $t$: Possible
singularities in \eqref{ret<0} and \eqref{ret>0} are at
$2t+4\ln\frac{f}{4}+2\pi i m=0$ and at  $-2t+4\ln\frac{f}{4}+2\pi i
m=0$, respectively. However, since $\ln\frac{f}{4}<0$, both cases are
excluded by the corresponding constraints $\mathrm{Re} \, t \, <0$ and
$\mathrm{Re} \, t \, >0$. We therefore conclude our search for
 the singularities with the conclusion that for the quasi-static
 state, the singularities in the complex $t$ plane for are absent.

We leave an analysis beyond the quasi-static approximation for further
work. Our result for the quasi-static case is very different from a
physical state undergoing thermalization, as described by the moving shell.
The shell shows distinct trajectories in Kruskal coordinate for the
quasi-static and moving shell cases: In the quasi-static state, the shell ``moves'' along $x_K^2-t_K^2=\text{constant}$, while the moving shell takes a trajectory $\frac{dx_K}{dt_K}=-1$. The difference is analogous to that of a free-falling object and a standing, but Unruh accelerating object \cite{unruh}. It remains to be seen whether singularities emerge in the complex $t$ plane for a physical thermalizing state. 

\section{Discussion}

We have generalized the divergence matching method for a gravitational collapse model. Application of this method leads to predictions on the evolution of the singularities in unequal time retarded correlator $G^R(t,t')=\theta(t-t')[O(t),O(t')]$ for a state in thermalization process. The singularities of $t$ are dictated by the geodesics of a bouncing light ray initially starting at $t'$. The fact that we found singularities at real $t$ can be traced to the nature of the divergence matching method, which is essentially a WKB approximation for real $\omg$. We have shown with an example of quasi-static state that possible singularities may exist also in the complex $t$ plane. It will be very interesting to confirm this, possibly by extending the current WKB approximation to complex $\omg$. The singularities at real $t$ coming from the normal modes have clear separation between contributions from positive and negative frequencies as shown in \eqref{Gnr_omg}. This will not be the case for singularities at complex $t$ coming from the QNM, in which no $i\eps$ prescription is needed. This is related to the fact that in the positive and negative frequency in the black hole background is defined with respect to the Kruskal time. It will be interesting to see if the evolution of singularities at complex $t$ can shed some light on this.

Another interesting point to note is that the formulation of
Section.\ref{sec_qnm} does not require $|\omg|\gg1$. Actually it is
sufficient to have $\frac{|\omg|}{\sqrt{f}}\gg 1$, which is always
realized when the shell is sufficient close to the horizon. However
there is one exception: the gapless hydrodynamic mode. For the case of
the dilaton, the QNM does not contain a gapless mode. However a gapless hydrodynamic mode does become relevant for example for a probe gauge or graviton field, due to the presence of conserved charges \cite{kovtun}. It will be interesting to study the evolution of the hydrodynamic mode. We hope to return to these issues in the future.
\\\\
\noindent{\it Note Added}: When the paper is near completion, we received \cite{vuorinen}, which has partial overlap with Section 5 of our paper. Qualitative agreement has been found between the QNM that are asymptotically real (normal modes in our paper and blue points in their plot). The counterpart of their red point seems to be absent in our paper. The particular QNM may be related to the hydrodynamic mode unique in their model.

\section*{Acknowledgements}

We would like to thank C.~Hoyos, S.~Stricker and A.~Vuorinen for useful discussions.
S.~L.~is supported by the Alexander von Humboldt Foundation. 
This work has been supported in part by the `Excellence
Cluster for Fundamental Physics: Origin and Structure of the
Universe'.

\appendix

\section{Residue at the QNM}
We closely follow section 3.3 of \cite{NS} in the calculation of the residue. Readers are encouraged to refer to \cite{NS} for more details. The tortoise coordinate is defined through $x=\int_0^r\frac{dr'}{f(r')}$, with an appropriate choice of branch cut. Near the boundary $r=\infty$, the solution to the wave equation is given by
\begin{align}\label{wkb_bdry}
\Phi(x)\sim &C_+\sqrt{2\pi\omg(x-x_0)}J_{\frac{j_\infty}{2}}(\omg(x-x_0))+C_-\sqrt{2\pi\omg(x-x_0)}J_{\frac{-j_\infty}{2}}(\omg(x-x_0)) \no
&\sim \(C_+e^{i\beta_+}+C_-e^{i\beta_-}\)e^{i\omg(x-x_0)}+\(C_+e^{-i\beta_+}+C_-e^{-i\beta_-}\)e^{-i\omg(x-x_0)} ,
\end{align}
where $\beta_\pm=\frac{\pi}{4}(1\pm j_\infty)$. Near the singularity $r=0$, the solution takes the following form
\begin{align}\label{wkb_center}
\Phi(x)&\sim B_+\sqrt{2\pi\omg x}J_{\frac{j}{2}}(\omg x)+B_-\sqrt{2\pi\omg x}J_{-\frac{j}{2}}(\omg x) \no
&\sim \(B_+e^{-i\alpha_+}+B_-e^{-i\alpha_-}\)e^{i\omg x}+\(B_+e^{i\alpha_+}+B_-e^{i\alpha_-}\)e^{-i\omg x},
\end{align}
where $\alpha_\pm=\frac{\pi}{4}(1\pm j)$.
\eqref{wkb_bdry} and \eqref{wkb_center} can be matched along the same Stokes line. To impose the ingoing boundary condition at the horizon, which lies on another Stokes line, we need to rotate \eqref{wkb_center} to the same Stokes line as the horizon. After the rotation, \eqref{wkb_center} becomes
\begin{align}
\Phi(x)\sim \(B_+e^{-i\alpha_+}+B_-e^{-i\alpha_-}\)e^{i\omg x}+\(B_+e^{-3i\alpha_+}+B_-e^{-3i\alpha_-}\)e^{-i\omg x}.
\end{align}
The ingoing boundary condition gives
\begin{align}\label{horizon_bc}
B_+e^{-3i\alpha_+}+B_-e^{-3i\alpha_-}=0.
\end{align}
Matching \eqref{wkb_bdry} and \eqref{wkb_center} and using \eqref{horizon_bc}, we obtain
\begin{align}\label{C_ratio}
\frac{C_+}{C_-}=\frac{\(e^{2i\alpha_+}-e^{2i\alpha_-}\)e^{i\omg x-i\beta_-}-\(e^{4i\alpha_+}-e^{4i\alpha_-}\)e^{-i\omg x+i\beta_-}}{-\(e^{2i\alpha_+}-e^{2i\alpha_-}\)e^{i\omg x-i\beta_+}+\(e^{4i\alpha_+}-e^{4i\alpha_-}\)e^{-i\omg x+i\beta_+}}.
\end{align}
For AdS$_d$ Schwarzschild, $j=0$ and $j_\infty=d-1$. \eqref{C_ratio} is undefined as both the denominator and the numerator vanish as $j=0$. We should use the L'Hopital rule to obtain
\begin{align}
\frac{C_+}{C_-}=\frac{e^{i\omg x-i\beta_-}-2ie^{-i\omg x+i\beta_-}}{-e^{i\omg x-i\beta_+}+2ie^{-i\omg x+i\beta_+}}.
\end{align}
The vanishing of the denominator gives the locations of the QNM at $\omg_n=\frac{n\pi+\theta}{x_0}$, with $\theta=\beta_++\frac{\ln 2i}{2i}$. The residue is also easily obtained as
\begin{align}
res\(\frac{C_+}{C_-}\)=-\frac{\sin\frac{\pi(d-1)}{2}}{x_0}.
\end{align}
Expanding \eqref{wkb_bdry} as $x\to x_0$($r\to\infty$), we find the residue of the retarded correlator at QNM $\omg_n$ given by
\begin{align}
res(\omg=\omg_n)=\frac{\Gm(\frac{3-d}{2})}{\Gm(\frac{d+1}{2})}\(-\frac{\omg_n}{2}\)^{d-1}res\(\frac{C_+}{C_-}\),
\end{align}
which in the limit $d\to 5$ reduces to $-\frac{\pi\omg_n^4}{32x_0}$.


\begin{thebibliography}{99}

\bibitem{minwalla}
  S.~Bhattacharyya and S.~Minwalla,
  JHEP {\bf 0909} (2009) 034
  [arXiv:0904.0464 [hep-th]].

\bibitem{CY}
  P.~M.~Chesler, L.~G.~Yaffe,
  Phys.\ Rev.\ Lett.\  {\bf 102 } (2009)  211601.
  [arXiv:0812.2053 [hep-th]].\\
  P.~M.~Chesler, L.~G.~Yaffe,
  Phys.\ Rev.\  {\bf D82 } (2010)  026006.
  [arXiv:0906.4426 [hep-th]].\\
  P.~M.~Chesler, L.~G.~Yaffe,
  Phys.\ Rev.\ Lett.\  {\bf 106 } (2011)  021601.
  [arXiv:1011.3562 [hep-th]].

\bibitem{bizon}
  P.~Bizon and A.~Rostworowski,
  Phys.\ Rev.\ Lett.\  {\bf 107} (2011) 031102
  [arXiv:1104.3702 [gr-qc]].\\
  O.~J.~C.~Dias, G.~T.~Horowitz and J.~E.~Santos,
  arXiv:1109.1825 [hep-th].

\bibitem{zayas}
  D.~Garfinkle, L.~A.~Pando Zayas,
  Phys.\ Rev.\  {\bf D84 } (2011)  066006.
  [arXiv:1106.2339 [hep-th]].\\
  D.~Garfinkle, L.~A.~Pando Zayas and D.~Reichmann,
  JHEP {\bf 1202} (2012) 119
  [arXiv:1110.5823 [hep-th]].

\bibitem{heller}
  M.~P.~Heller, R.~A.~Janik and P.~Witaszczyk,
  arXiv:1103.3452 [hep-th].\\
  M.~P.~Heller, R.~A.~Janik and P.~Witaszczyk,
  arXiv:1203.0755 [hep-th].

\bibitem{gubser}
  H.~Bantilan, F.~Pretorius and S.~S.~Gubser,
  arXiv:1201.2132 [hep-th].

\bibitem{star}
  X.~Arsiwalla, J.~de Boer, K.~Papadodimas and E.~Verlinde,
  JHEP {\bf 1101} (2011) 144
  [arXiv:1010.5784 [hep-th]].

\bibitem{uppsala}
  U.~H.~Danielsson, E.~Keski-Vakkuri, M.~Kruczenski,
  Nucl.\ Phys.\  {\bf B563 } (1999)  279-292.
  [hep-th/9905227].\\
  U.~H.~Danielsson, E.~Keski-Vakkuri, M.~Kruczenski,
  JHEP {\bf 0002 } (2000)  039.
  [hep-th/9912209].

\bibitem{giddings}
  S.~B.~Giddings, A.~Nudelman,
  JHEP {\bf 0202 } (2002)  003.
  [hep-th/0112099].

\bibitem{HLR}
  V.~EHubeny, H.~Liu, M.~Rangamani,
  JHEP {\bf 0701 } (2007)  009.
  [hep-th/0610041].

\bibitem{shell}
  S.~Lin, E.~Shuryak,
  Phys.\ Rev.\  {\bf D78 } (2008)  125018.
  [arXiv:0808.0910 [hep-th]].

\bibitem{kovchegov}
  H.~R.~Grigoryan, Y.~V.~Kovchegov,
  JHEP {\bf 1104 } (2011)  010.
  [arXiv:1012.5431 [hep-th]].

\bibitem{ELN}
  J.~Erdmenger, S.~Lin and T.~H.~Ngo,
  JHEP {\bf 1104} (2011) 035
  [arXiv:1101.5505 [hep-th]].

\bibitem{EHL}
  J.~Erdmenger, C.~Hoyos and S.~Lin,
  JHEP {\bf 1203} (2012) 085
  [arXiv:1112.1963 [hep-th]].

\bibitem{11author}
  V.~Balasubramanian, A.~Bernamonti, J.~de Boer, N.~Copland, B.~Craps, E.~Keski-Vakkuri, B.~Muller, A.~Schafer {\it et al.},
  Phys.\ Rev.\ Lett.\  {\bf 106 } (2011)  191601.
  [arXiv:1012.4753 [hep-th]].\\
  V.~Balasubramanian, A.~Bernamonti, J.~de Boer, N.~Copland, B.~Craps, E.~Keski-Vakkuri, B.~Muller, A.~Schafer {\it et al.},
  Phys.\ Rev.\  {\bf D84 } (2011)  026010.
  [arXiv:1103.2683 [hep-th]].

\bibitem{teaney}
  S.~Caron-Huot, P.~M.~Chesler, D.~Teaney,
  Phys.\ Rev.\  {\bf D84 } (2011)  026012.
  [arXiv:1102.1073 [hep-th]].

\bibitem{headrick}
  H.~Ebrahim, M.~Headrick,
  [arXiv:1010.5443 [hep-th]].

\bibitem{lopez}
  J.~Abajo-Arrastia, J.~Aparicio, E.~Lopez,
  JHEP {\bf 1011 } (2010)  149.
  [arXiv:1006.4090 [hep-th]].\\
  J.~Aparicio, E.~Lopez,
  [arXiv:1109.3571 [hep-th]].

\bibitem{johnson}
  T.~Albash, C.~V.~Johnson,
  New J.\ Phys.\  {\bf 13 } (2011)  045017.
  [arXiv:1008.3027 [hep-th]].

\bibitem{takayanagi}
  T.~Takayanagi and T.~Ugajin,
  JHEP {\bf 1011} (2010) 054
  [arXiv:1008.3439 [hep-th]].

\bibitem{teaney2}
  P.~M.~Chesler and D.~Teaney,
  arXiv:1112.6196 [hep-th].

\bibitem{hoyos}
  I.~Amado and C.~Hoyos-Badajoz,
  JHEP {\bf 0809} (2008) 118
  [arXiv:0807.2337 [hep-th]].

\bibitem{israel}
  W.~Israel,
  Nuovo Cim.\ B {\bf 44S10} (1966) 1
   [Erratum-ibid.\ B {\bf 48} (1967) 463]
   [Nuovo Cim.\ B {\bf 44} (1966) 1].

\bibitem{svrjhep}
  K.~Skenderis and B.~C.~van Rees,
  JHEP {\bf 0905} (2009) 085
  [arXiv:0812.2909 [hep-th]].

\bibitem{hubeny}
  L.~Fidkowski, V.~Hubeny, M.~Kleban and S.~Shenker,
  JHEP {\bf 0402} (2004) 014
  [hep-th/0306170].

\bibitem{NS}
  J.~Natario and R.~Schiappa,
  Adv.\ Theor.\ Math.\ Phys.\  {\bf 8} (2004) 1001
  [hep-th/0411267].

\bibitem{festuccia}
  G.~Festuccia and H.~Liu,
  JHEP {\bf 0604} (2006) 044
  [hep-th/0506202].

\bibitem{starinets}
  E.~Berti, V.~Cardoso and A.~O.~Starinets,
  Class.\ Quant.\ Grav.\  {\bf 26} (2009) 163001
  [arXiv:0905.2975 [gr-qc]].

\bibitem{unruh}
  W.~G.~Unruh,
  Phys.\ Rev.\ D {\bf 14} (1976) 870.

\bibitem{kovtun}
  P.~K.~Kovtun and A.~O.~Starinets,
  Phys.\ Rev.\ D {\bf 72} (2005) 086009
  [hep-th/0506184].

\bibitem{vuorinen}
  R.~Baier, S.~A.~Stricker, O.~Taanila and A.~Vuorinen,
  arXiv:1205.2998 [hep-ph].

\end{thebibliography}
\end{document}